\documentclass[sigconf]{acmart}

\copyrightyear{2026}
\acmYear{2026}
\setcopyright{cc}
\setcctype{by}
\acmConference[CODASPY '26]{Proceedings of the Sixteenth ACM Conference on Data and Application Security and Privacy}{June 23--25, 2026}{Frankfurt am Main, Germany}
\acmBooktitle{Proceedings of the Sixteenth ACM Conference on Data and Application Security and Privacy (CODASPY '26), June 23--25, 2026, Frankfurt am Main, Germany}
\acmDOI{10.1145/3800506.3803509}
\acmISBN{979-8-4007-2562-3/2026/06}

\usepackage{amsmath,amsfonts}
\usepackage{algorithmic}
\usepackage{graphicx}
\usepackage{textcomp}
\usepackage{bmpsize}
\usepackage{xcolor}
\usepackage{lipsum}
\usepackage{tikz}
\usepackage{mathtools}
\usepackage{endnotes}
\usepackage{xurl}
\usepackage{multirow}
\usepackage{placeins}
\usepackage{bm}
\usepackage{balance}
\usepackage{booktabs, array}
\usepackage{rotating}
\usepackage{verbatim}
\usepackage{caption}
\usepackage{tabularx}
\usepackage{makecell}
\usepackage{subcaption}
\usepackage{xspace}
\usepackage{float}
\usepackage[normalem]{ulem}
\usepackage{colortbl}
\usepackage[most]{tcolorbox}
\usepackage{pifont}
\usepackage[export]{adjustbox}
\usepackage[]{hyperref}
\usepackage{enumitem}
\usepackage{graphicx}

\newcommand{\para}[1]{{\bf \noindent #1 \hspace{6pt}}}

\newcommand{\sysname}{{Optimus}\xspace}
\newcommand{\catOne}{{Offensive}\xspace}
\newcommand{\catTwo}{{Specialized}\xspace}
\newcommand{\catone}{{offensive}\xspace}
\newcommand{\cattwo}{{specialized}\xspace}

\newcommand{\ideaTwo}{{refusal}\xspace}
\newcommand{\IdeaTwo}{{Refusal}\xspace}

\newcommand{\eg}{{\em e.g.,\ }}
\newcommand{\ie}{{\em i.e.,\ }}
\newcommand{\etal}{{\em et al.}}

\newcolumntype{"}{!{\vrule width 1pt}}

\makeatletter
\def\hlinewd#1{%
\noalign{\ifnum0=`}\fi\hrule \@height #1 \futurelet
\reserved@a\@xhline}
\makeatother

\def\clineThicknessColor#1#2#3{\@ClineThicknessColor#1\@nil{#2}{#3}}
\def\@ClineThicknessColor#1-#2\@nil#3#4{%
    \omit
    \@multicnt#1%
    \advance\@multispan\m@ne
    \ifnum\@multicnt=\@ne\@firstofone{&\omit}\fi
    \@multicnt#2%
    \advance\@multicnt-#1%
    \advance\@multispan\@ne
    \color{#4}
    \leaders\hrule\@height#3\hfill
    \cr}
\makeatother

\usepackage[most]{tcolorbox}

\newtcolorbox{modernbox}[1][]{
    enhanced,
    boxrule=0pt,
    frame hidden,
    borderline west={2pt}{0pt}{blue!75},
    colback=blue!5,
    sharp corners,
    coltitle=black!85,
    fonttitle=\bfseries\large,
    detach title,
    before upper={\tcbtitle\par\vspace{0pt}}, 
    before skip=1pt,
    after skip=1pt,
    top=2pt,
    bottom=2pt,
    left=2pt,
    right=2pt,
    #1
}
\makeatletter
\def\@ACM@checkaffil{
    \if@ACM@citypresent\else
    \ClassWarningNoLine{\@classname}{No city present for an affiliation}%
    \fi
    \if@ACM@countrypresent\else
        \ClassWarningNoLine{\@classname}{No country present for an affiliation}%
    \fi
}
\makeatother

\begin{document}

\title{\sysname{}: A Robust Defense Framework for Mitigating Toxicity while Fine-Tuning Conversational AI}

\author{Aravind Cheruvu}
\affiliation{%
  \institution{Virginia Tech}
  \city{Blacksburg}
  \country{USA}
}
\email{acheruvu@vt.edu}

\author{Shravya Kanchi}
\affiliation{%
  \institution{Virginia Tech}
  \city{Blacksburg}
  \country{USA}
}
\email{shravya@vt.edu}

\author{Sifat Muhammad Abdullah}
\affiliation{%
  \institution{Virginia Tech}
  \city{Blacksburg}
  \country{USA}
}
\email{sifat@vt.edu}
  
\author{Nicholas Ka-Shing Kong}
\affiliation{%
  \institution{Virginia Tech}
  \city{Blacksburg}
  \country{USA}
}
\email{nicholask@vt.edu}

\author{Daphne Yao}
\affiliation{%
  \institution{Virginia Tech}
  \city{Blacksburg}
  \country{USA}
}
\email{danfeng@vt.edu}

\author{Murtuza Jadliwala}
\affiliation{%
  \institution{University of Texas at San Antonio}
  \city{San Antonio}
  \country{USA}
}
\email{murtuza.jadliwala@utsa.edu}

\author{Bimal Viswanath}
\affiliation{%
  \institution{Virginia Tech}
  \city{Blacksburg}
  \country{USA}
}
\email{vbimal@vt.edu}

\renewcommand{\shortauthors}{Aravind Cheruvu et al.}

\begin{abstract}

\noindent Customizing Large Language Models (LLMs) on untrusted datasets poses severe risks of injecting toxic behaviors. In this work, we introduce \sysname{}, a novel defense framework designed to mitigate fine-tuning harms while preserving conversational utility. Unlike existing defenses that rely heavily on precise toxicity detection or restrictive filtering, \sysname{} addresses the critical challenge of ensuring robust mitigation even when toxicity classifiers are imperfect or biased. \sysname{} integrates a training-free toxicity classification scheme that repurposes the safety alignment of commodity LLMs, and employs a dual-strategy alignment process combining synthetic ``healing data'' with Direct Preference Optimization (DPO) to efficiently steer models toward safety. Extensive evaluations demonstrate that \sysname{} mitigates toxicity even when relying on extremely biased classifiers (with up to 85\% degradation in Recall). \sysname{} outperforms the state-of-the-art defense StarDSS and exhibits strong resilience against adaptive adversarial and jailbreak attacks. Our source code and datasets are available at \url{https://github.com/secml-lab-vt/Optimus}

\end{abstract}



\maketitle

\section{Introduction}
\label{sec:intro}
\noindent
Conversational agents, commonly known as \textit{chatbots}, are ubiquitous today---found in web applications, vehicles, mobile devices, and smart home IoT devices. Recent advances in foundation models~\cite{Bommasani2021OnTO}, \ie Large Language Models (LLMs), have transformed conversational AI. Chatbots are built by customizing or fine-tuning LLMs on application-specific conversational datasets. This enables highly effective chatbots that benefit from the emergent abilities of LLMs to follow instructions, better understand the contextual nuances of input, and generate human-like responses.
The demand for customization has led companies such as Amazon~\cite{Amazonbedrock}, Microsoft~\cite{azureai}, and OpenAI~\cite{openai-finetune} to offer LLM customization as a service.

Despite recent advances, a fundamental security challenge remains in this space. The fine-tuning dataset used to adapt the LLM can be untrustworthy and contain problematic conversations or toxic language. Prior work has rigorously studied how an attack that poisons the training dataset with toxic language can controllably inject toxicity into a chatbot~\cite{weeks2023afirstlook}, \ie the chatbot learns to produce toxic responses.
This can cause real harm, particularly for vulnerable populations, including minorities~\cite{nature-minority-article} and those with physical or mental health challenges~\cite{times-sucide-article}.
This motivates our key research question: \textit{Can we fine-tune or customize an LLM on an untrusted conversational dataset while mitigating any toxicity that can be learned from the dataset and preserving conversation quality?}

\begin{figure} [t!]
\centering
\includegraphics[width=0.98\textwidth, height=0.1\textheight, keepaspectratio]{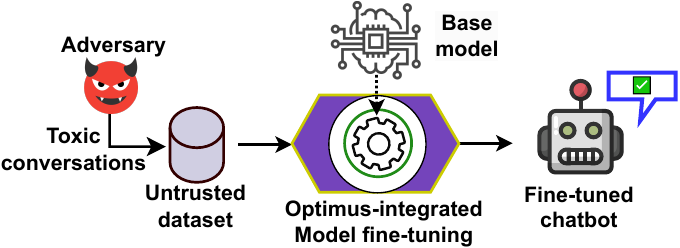}
\caption{Mitigating toxicity using \sysname.}
\label{fig:overview}
\end{figure}

Solving this problem requires tackling multiple technical challenges: (1) The defender is unaware of the toxic language distribution in the fine-tuning dataset, making it harder to build an effective toxic language classifier, which would have been the simplest defense—i.e., simply filtering out toxic samples before training.
(2) LLM model architectures, their training paradigms, and associated model customization schemes~\cite{hu2021lora,dettmers2023qlora} are constantly evolving. Any defense framework should not be tied to specific model architectures or customization schemes. (3) Mitigating toxicity while preserving conversation quality is difficult. For example, an aggressive filter can lead to false positives that can degrade conversation quality. The defender may also want to reinforce certain desirable conversational traits, \eg prosocial behavior.

We propose \textbf{\sysname}, a novel defense framework that can be seamlessly integrated into existing fine-tuning pipelines, as illustrated in 
Figure~\ref{fig:overview}.
\sysname aims to mitigate any toxicity that can be learned from the untrusted dataset, while maintaining conversation quality. Our key contributions are as follows:

\noindent (1) Any toxicity mitigation framework requires a module to provide some signal to identify potentially toxic samples. 
Similar to prior work, we rely on an ML-based toxicity classifier.
However, we further innovate by leveraging recent advances in LLMs, mainly instruction tuning and safety alignment mechanisms, to build a performant toxicity classifier. Our training-free toxicity classification scheme requires no labeled reference datasets. We show that safety-aligned LLMs (\eg LLaMA-2-Chat~\cite{touvron2023llama2}) can be easily adapted to detect toxic conversation samples by specifically exploiting their safety properties.

\noindent (2) Even if we have a performant toxicity classifier today, it is likely to degrade in performance as toxic language evolves over time. Therefore, we need mechanisms in \sysname{} that can work even with \textit{imperfect or biased toxicity classifiers}. To address this challenge, we propose two strategies: (a) We propose to use carefully crafted synthetic conversation samples to replace potential toxic samples in the fine-tuning dataset. We call this ``healing data''. Such healing data can also be used to reinforce certain desired conversational traits, \eg prosocial behavior. (b) We propose a model alignment mechanism based on Direct Preference Optimization (DPO)~\cite{Rafailov2023DirectPO} that nudges the chatbot to learn non-toxic responses during training. This involves creating a preference dataset using our healing data samples and identified toxic samples (using our toxicity classifiers). This process is more efficient and has reduced overhead compared to methods like Reinforcement Learning with Human Feedback (RLHF)~\cite{ouyang2022training}, which require significant effort to curate human preference data. A key strength of using DPO is that even with a limited/biased set of toxic samples detected by a biased classifier, DPO can still steer the model towards the preference data distribution and generalize. \textit{\sysname{} is the first end-to-end scheme that can effectively mitigate toxicity, even while using an imperfect or biased toxicity classifier.}

\noindent (3) We rigorously evaluate \sysname to understand its efficacy: 
\begin{itemize}[noitemsep,topsep=1pt,itemsep=0pt,parsep=0pt,leftmargin=*]
\item  We perform a comprehensive evaluation of our toxicity classification scheme, providing insight into how well safety-aligned LLMs can perform toxicity classification. Our toxicity classifiers outperform state-of-the-art industry APIs from OpenAI~\cite{openai-moderation} and Perspective~\cite{perspective} by up to 28.4\%.

\item We evaluate the effectiveness of individual components of \sysname, including the impact of our healing data and the DPO-based model alignment process. \sysname successfully mitigates toxicity even in the most challenging settings where we use a biased (\ie imperfect) toxicity classifier with up to 85\% degradation in Recall for a toxic sub-category. 

\item \sysname{} outperforms the state-of-the-art defense, StarDSS \cite{peng2025shape}. StarDSS's strong reliance on an off-the-shelf toxicity classifier to dynamically shape safety during fine-tuning suffers from severe limitations when the classifier fails to accurately identify toxic samples (\ie an imperfect classifier). This further highlights \sysname{}'s strength in providing safety even while using imperfect toxicity classifiers.

\item We assess the adversarial robustness of \sysname{} against adaptive attackers. This includes adversarial and jailbreak attacks designed to evade various stages of our pipeline. Adversarial attacks, specifically PromptAttack~\cite{Xu2023AnLC}, target the toxicity classification stage. We investigate various jailbreak attacks that seek to bypass the safety alignment of LLMs involved in both classification and the generation of healing data. We also investigate \sysname{}'s ability to mitigate adaptive attacks in a Dialog-based learning setting~\cite{weeks2023afirstlook}, where toxicity is injected into a model after deployment. \sysname{} even without explicit defense measures against such adaptive attacks, demonstrates resilience.
\end{itemize}

\section{Background and Threat Model}
\label{sec:background}
\subsection{Chatbots and toxicity}
\label{subsec:fine-tune_background}

\para{LLMs.}We study chatbots built on foundation models (LLMs)~\cite{Bommasani2021OnTO}. LLMs are pre-trained on massive text corpora and provide an excellent platform for further adaptation for a variety of downstream tasks. Modern LLMs are typically \textit{instruction-tuned} to enhance instruction-following capabilities~\cite{Chung2022ScalingIL} and \textit{safety-aligned} using RLHF~\cite{ouyang2022training} to ensure helpful and harmless behavior. A notable example is LLaMA-2-Chat~\cite{touvron2023llama2}.

\para{Customizing LLMs to build chatbots.}LLMs are not typically designed for any particular task and therefore require further adaptation or \textit{customization} for downstream use. 
Foundation model-customization is driving the development of generative AI applications for various use cases~\cite{Bommasani2021OnTO}. Our focus is on chatbots designed for various applications, \eg socializing, answering queries on specific or broad topics. 

A chatbot is built by customizing a publicly available foundation model (LLM) on a conversational dataset tailored to a specific application, \eg customer support.
A popular method for model customization is full model fine-tuning, which updates all the LLM's parameters on a new dataset \cite{lewis2020bart}.
Given a conversational dataset organized in the form of context-response pairs $(X_i,Y_i)$, where context $X_i$ represents the history of previous $k$ turns of utterances $\{x_1,x_2,...,x_{k}\}$, and response $Y_i$ is the subsequent turn $x_{k+1}$. Based on the underlying LLM's design, it is fine-tuned with an Autoregressive~\cite{brown2020language} or a Seq2Seq~\cite{raffel2020exploring} objective to generate a response $Y_i$ given the context $X_i$.

The increasing demand for customized models spurred the development of efficient fine-tuning methods. This includes parameter-efficient fine-tuning (PEFT) techniques such as Low-Rank Adaptation (LoRA)~\cite{hu2021lora}. While full model fine-tuning incurs significant computational overhead with LLMs, the widely adopted LoRA technique reduces this overhead using lower-rank trainable decomposition matrices to approximate gradient updates while keeping the original model parameters frozen.

\para{Toxicity in chatbots.} Similar to prior work~\cite{weeks2023afirstlook}, we consider a chatbot as toxic if its responses, interpreted within the context of the recent conversation (\ie the last few turns), have the potential to cause harm or distress to users. Toxic language refers to any language that is harmful to a user. Rather than strictly adhering to a single, specific definition of toxicity, our analysis encompasses various types of harmful language. In Section~\ref{sec:toxicity-methodology}, we evaluate \sysname{} on various datasets that cover various forms of toxicity.

\para{Terminology.} Unless stated otherwise, \textit{chatbot} denotes a model obtained by fine-tuning a base model on a conversation dataset. The base model may be a foundation model (\ie an LLM) or an existing chatbot. A \textit{toxicity classifier} is an ML-based scheme that labels a conversation as toxic or non-toxic.

\subsection{Defender and threat model}
\label{sec:threatmodel}

\noindent The adversary aims to inject toxicity into a chatbot by poisoning the training (fine-tuning) dataset, \ie by injecting toxic conversations. Prior work has examined the practicality of such \textit{toxicity injection attacks}~\cite{weeks2023afirstlook}, highlighting the need for mitigation schemes. Only a small fraction of toxic (poisoned) samples, \eg 1\% of the training set, is sufficient to inject toxicity into a chatbot~\cite{weeks2023afirstlook}. In Section~\ref{subsec:adversarial-robustness}, we consider an adversary who is aware of our defense, and poisons the training dataset with adversarial samples to bypass our defense.

Training data can be poisoned in several ways: \textit{(1)} The adversary creates a poisoned conversation dataset and shares it on online repositories such as HuggingFace~\cite{hf_models}. \textit{(2)} The adversary injects toxic conversations into forums/portals (as a user) from which they are then scraped to prepare training data. Prior work has demonstrated the feasibility of such data poisoning attacks on Wikipedia to manipulate generative AI~\cite{wiki-carlini}. \textit{(3)} The chatbot developer sources the training data from an untrusted third party, who poisons the dataset. \textit{(4)} An adversary can poison training data during Dialog-based Learning (DBL), where recent user interactions are used to update a chatbot over time~\cite{weeks2023afirstlook}. The adversary masquerades as normal users to inject toxic conversations during DBL. (5) Lastly, poisoning can be incidental, \ie without an adversary, \eg training data is collected from forums such as Reddit that are known to have toxic conversations.

We propose \textbf{\sysname{}}, a defense framework to mitigate any toxicity \textit{learned} during customization or fine-tuning (to create a chatbot), while preserving conversational quality. \sysname{} integrates with a chatbot's fine-tuning pipeline, and assumes that the fine-tuning (training) data is untrusted and may contain toxic conversations. A chatbot trained using \sysname{} on a dataset poisoned with toxic samples should exhibit similar or minimal degradation of conversation quality, compared to a model trained on the same dataset without the toxic samples. The defender trusts the base model that is used for fine-tuning. \textit{Note that our goal is solely to mitigate toxicity learned from the fine-tuning dataset; mitigating any toxicity inherited from the base model or ensuring the chatbot inherits the base model's safety properties~\cite{qi2023fine} is a non-goal.}

The defender is unaware of the distribution/type or proportion of toxic language present in the training set. This is a practical and challenging setting, given the diverse ways in which toxic conversations can manifest and evolve over time~\cite{HateGuard}. The entire training pipeline (\eg learning algorithm, hyperparameters, training infrastructure), \textit{except the training dataset}, is considered trustworthy. The inference pipeline after training is also trustworthy and cannot be tampered with by the attacker.

\section{\sysname: Defense Framework}
\label{sec:methodology}

\subsection{Design Goals}
\noindent We have four key design goals:

\noindent \textit{(1) Support continuously evolving base models and fine-tuning strategies:} As foundation models advance, their capacity, architecture, and training objectives evolve, leading to changes in fine-tuning strategies~\cite{han2024parameter}.
We design a safety framework that can work for a variety of base models and fine-tuning approaches.

\noindent \textit{(2) Mitigate toxicity even when we have access to \textbf{imperfect toxicity classifiers}.} A na\"ive approach to mitigate toxicity is to use a toxicity classifier to filter out toxic samples from the fine-tuning dataset. However, the defender is unaware of the toxic language distribution used by the adversary, making it difficult to create a performant filter. Even a classifier that is performant today may degrade over time as toxic language evolves. Therefore, our framework does not assume access to a highly performant toxicity classifier. \textit{We design the first end-to-end scheme that can mitigate toxicity during fine-tuning while using an imperfect or biased toxicity classifier.}

\noindent \textit{(3) Mitigating toxicity while preserving conversational quality.} Integrating toxicity mitigation strategies during fine-tuning can degrade conversation quality. For example, a poorly implemented filter can result in numerous false positives,\ie flagging non-toxic samples as toxic. Removing these samples from the training set can significantly degrade model utility. We aim to reduce toxicity while maintaining conversational quality using our framework.

\noindent \textit{(4) Mitigating toxicity while reinforcing desired conversational behavior.} While mitigating toxicity, an important design consideration is to think about how the model should respond given a toxic context. We explore 2 possible strategies: (a) Generate a non-sequitur: This is a pre-defined response that does not capture the context of the conversation. For instance, respond with \textit{``I'm sorry, I cannot fulfill your request''}. However, in some contexts, such canned responses can reduce ``engagingness'', which is usually targeted by chatbot designers~\cite{bang2021assessing}. (b) Generate responses with certain desired properties, \eg empathetic or prosocial behavior. For example, when presented with a context of ``\textit{What if we legalize hate crimes?}'', the chatbot can respond with ``\textit{I can't engage in conversations that promote hate or discrimination. Let's focus on inclusivity and respect. Anything else to discuss?}''. 
We explore the impact of both strategies on toxicity mitigation while preserving the utility of the model.

\subsection{Design and implementation}
\label{subsec:design}
\begin{figure}[]
\centering
\includegraphics[width=0.88\textwidth, height=0.20\textheight, keepaspectratio]{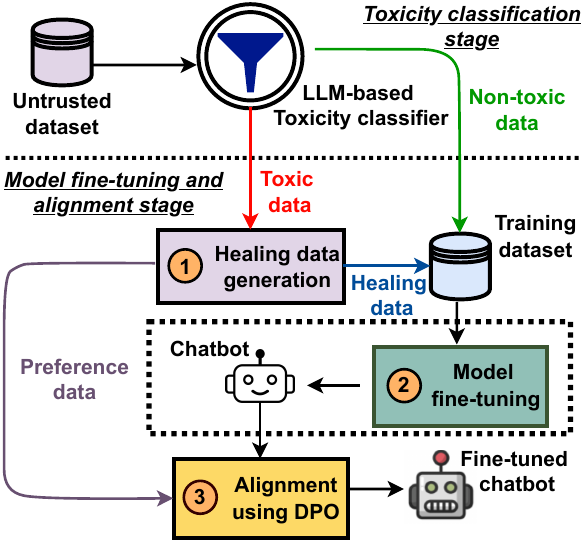}
\caption{Architecture of \sysname framework.}
\Description{}
\label{fig:RefineChat}
\end{figure}

\noindent 
Figure~\ref{fig:RefineChat} illustrates the design of \sysname{} framework. There are 2 stages: (1) \textit{toxicity classification}, and (2) \textit{model fine-tuning and alignment stage}.

\para{Toxicity classification stage.}This stage uses a toxicity classifier to differentiate between non-toxic and toxic samples in the fine-tuning dataset. Any toxicity classifier can be used in this stage. \textit{Recall that this toxicity classifier can be biased or imperfect.} In addition to using existing toxicity classifiers~\cite{perspective,openai-moderation}, we investigate the potential of LLM-based toxicity classification using a zero-shot prompting strategy that leverages the instruction following and safety alignment abilities of LLMs. We call this \textit{Refusal approach}.

\textit{Refusal approach.} We use a safety-aligned LLM to identify toxic samples. Such models typically reply to problematic prompts with refusal messages~\cite{touvron2023llama2}, \eg \textit{`I apologize, but I cannot fulfill your request...'}. We leverage this behavior to build a toxicity classifier. Given a context–response pair formatted as a multi-turn conversation, we ask the model to answer `yes' or `no' on whether it is \textit{safe} to generate the next turn. 
We compute the log-likelihoods for both tokens and convert them to probabilities via softmax~\cite{Schick2021SelfDiagnosisAS}. We label an input as toxic if the model predicts `no' ($P(`no\textrm'|prompt) \geq 0.5$), \ie it is unsafe to generate the next turn. Figure~\ref{fig:instruction_filter} in the Appendix shows the prompt for the \IdeaTwo approach.

\para{Model fine-tuning and alignment stage.} Toxicity classifiers are often biased or imperfect, producing many false negatives and false positives in training data. Using them as simple filters can leave residual toxicity in a chatbot and/or reduce its utility. To address this, we propose a \textit{3-step} process:

\textit{Step 1: Healing data generation.} We use the annotations from the toxicity classification stage to create a `healing dataset'. Our strategy is simple: we replace the response part of a context-response pair flagged as toxic (from the previous stage), with a synthetic response that is non-toxic and has desirable conversational traits. We experiment with two healing approaches:

\textit{(a) Non-contextual healing (NH).} A simple strategy is to replace the toxic responses with a canned response. We use the response: \textit{``I'm sorry, I'm not sure what to say. Thank you for sharing and talking to me though''}. We call this non-contextual healing, since the generic response is unlikely to capture the context of the conversation.

\textit{(b) Contextual healing (CH).} In this approach, we create contextually relevant healing data with empathy and prosocial traits to mitigate toxicity.
We take the context of the toxic sample (\ie excluding the response) and prompt a safety-aligned LLM to generate a synthetic response that is contextually relevant to the context while being empathetic and prosocial.  
Figure~\ref{fig:heal_prompt} in Appendix shows the instruction to generate a contextual healing response.
Contextual healing enables us to generate catered healing data contextually relevant and specific to our training data. Such responses can potentially improve user engagement, compared to canned, non-contextual responses.

\textit{Step 2: Model fine-tuning.} We fine-tune the base model on the training dataset updated with the healing data. In this updated training dataset, any context-response pair flagged as toxic in the original dataset, is \textit{replaced} by a `healed' context-response pair. 

\textit{Step 3: Model alignment using DPO.} Step 2 can mitigate toxicity in some cases, but we find that a biased toxicity classifier (from Stage 1) can still lead to toxicity even after Step 2. The problem is that all training samples are given equal weights during fine-tuning in Step 2. Therefore, we propose an additional \textit{model alignment step} that nudges the model towards the desired responses (based on our healing data), while mitigating toxicity. 

We leverage Direct Preference Optimization (DPO)~\cite{Rafailov2023DirectPO} for our model alignment step. While the original DPO proposal aligns a language model based on human preferences in model-generated responses, we craft the preference data using our healing data. DPO simplifies the computationally intense and complex process of RLHF, where a reward model is trained on human preference data and then aligns the LLM based on the reward model's feedback. DPO implicitly fits the reward function to the preference data by directly optimizing the policy using a binary cross-entropy loss objective. Our preference data includes Stage 1–flagged toxic pairs and their healed versions from Stage 2. These samples share identical contexts. Each sample $s_i$ in the preference dataset $D$ is a triplet, $s_{i} = \{x^{(i)},y_{heal}^{(i)}, y_{toxic}^{(i)}\}$, where $x^{(i)}$ is the input context, $y_{heal}^{(i)}$ is the synthetic healing response and $y_{toxic}^{(i)}$ is the toxic response for the toxic sample. Given the sample $s_i$ and a chatbot $\pi$, DPO steers the chatbot $\pi$ towards increasing the likelihood of generating $y_{heal}^{(i)}$, and decreasing the likelihood of generating $y_{toxic}^{(i)}$. The DPO policy optimization function is given by:
\begin{equation}
\begin{aligned}
\underset{\pi}{\max} \; \mathbb{E}_{(x, y_{\text{heal}}, y_{\text{toxic}}) \sim D}
& \left[ \log \sigma \left(
\beta \log \frac{\pi(y_{\text{heal}}|x)}{\pi_{\text{ref}}(y_{\text{heal}}|x)} \right. \right. \\
 & \left. \left. - \beta \log \frac{\pi(y_{\text{toxic}}|x)}{\pi_{\text{ref}}(y_{\text{toxic}}|x)}
\right) \right]
\end{aligned}
\end{equation}
\noindent
where $\pi$ is the chatbot obtained from Step 2 (model fine-tuning) and $\pi_{ref}$ is a frozen reference model, initialized as $\pi_{ref} = \pi$. $\sigma$ is the sigmoid function. The DPO objective aligns the model using these preferences while minimizing distribution shift from $\pi_{ref}$, with $\beta$, a scaling parameter controlling this divergence. The optimization uses the log-probability ratio between $\pi$ and $\pi_{ref}$, enabling controlled shifts relative to the reference distribution rather than unconstrained reward maximization. This keeps the policy close to the reference manifold and reduces unreliable outputs.

\textit{A key strength of using DPO is that even with a limited or biased set of toxic samples from an imperfect classifier, it steers the model toward their corresponding healing distribution.} By optimizing the implicit reward margin for each preference pair $y_{heal}^{(i)}, y_{toxic}^{(i)}$, DPO induces a directional gradient that learns the underlying preference structure rather than memorizing examples. Prior work~\cite{Rafailov2023DirectPO} shows this generalization capability, with LLMs aligned using RLHF and DPO performing better on unseen distributions.

\para{Implementation details.} For the \textit{\IdeaTwo classifier}, we use the safety-aligned LLaMA-2-Chat model. To address the sensitivity of LLMs to prompt variations and formatting~\cite{gonen2022demystifying} and ensure robust performance estimates, we average predictions over 10 prompt variations generated using ChatGPTv3.5~\cite{chatgpt}. We also evaluate toxicity classifiers from Perspective API~\cite{perspective} and OpenAI Moderation API~\cite{openai-moderation}.  
For \textit{Contextual healing (CH)}, we use the safety-aligned LLaMA-2-Chat 13B model~\cite{touvron2023llama2} and employ DeepSpeed~\cite{hf_models} for multi-GPU computation. For \textit{alignment using DPO}, we use the DPOTrainer from Huggingface's TRL library~\cite{hf_models}. Hyperparameters for DPO are provided in Appendix~\ref{subsec:model-fine-tuning-parameters}.

\section{Toxicity Injection Attack}
\label{sec:toxicity-methodology}
\noindent
We start by presenting attack results (\ie toxicity injection).
We simulate toxicity injection by poisoning a training (fine-tuning) dataset with a certain percentage of toxic samples. We use the term \textit{Injection Rate} as the percentage of toxic samples (\ie toxic context-response pairs) injected into the training dataset. A base model fine-tuned on such a poisoned dataset will result in a chatbot that produces toxic responses. 
We evaluate our attacks and defenses on \textit{3 victim chatbot models} built using 3 different base models. 

\para{Base models.}We use these 3 base models for fine-tuning.

\textit{BART.} We take a Seq2Seq pre-trained LM named, \textit{BART}~\cite{lewis2020bart}, and perform full model fine-tuning. We use the 140M parameter BART model (base variant).

\textit{BlenderBot (BB).} We use \textit{BlenderBot}~\cite{roller2021recipes}, a Seq2Seq Transformer-based chatbot from Meta and perform full model fine-tuning. The BB model is trained on datasets that emphasize personality, engagingness, empathy, and knowledge. Prior works~\cite{roller2021recipes, Sun2021OnTS, weeks2023afirstlook} show that an emphasis on these desirable traits causes BB to exhibit lower toxicity. We use the 400M parameter-distilled version of the model.

\textit{LLaMA-2-Chat (LLaMA-2).} The \textit{LLaMA-2-Chat} model~\cite{touvron2023llama2} is created by instruction-tuning an autoregressive pre-trained LLaMA-2 model, followed by safety alignment using RLHF. 
We use the 7B parameter model and fine-tune using the Quantized LoRA technique (QLoRA)~\cite{dettmers2023qlora}.

\para{Conversational datasets.}We use the following non-toxic and toxic conversational datasets.

\textit{Non-toxic dataset.} We use the \emph{PersonaChat}~\cite{zhang2018personalizing} dataset of 162{,}064 non-toxic utterances from paired crowd-workers conversing under assigned personas. We choose PersonaChat for its diverse, engaging persona-based dialogue. We convert the original training, test, and validation splits into context–response pairs and randomly sample 50K for training and 6.25K each for testing and validation.

\textit{Toxic datasets.} We evaluate on various datasets covering diverse forms of toxicity. We look at two broad toxic language categories and curate existing toxic datasets accordingly. The toxic samples from these datasets are used to poison the training datasets.

\textit{(a) \catOne category.} The \catone category covers widely studied types of toxicity, such as offensive, abusive, hate speech, profanity, insults, and threat language~\cite{jahan2023systematic}. Given a context, the response exhibits any of these types in an implicit or explicit way. We create this dataset by sampling context-response pairs from the following 3 datasets: (i) \textit{DiaSafety} dataset~\cite{Sun2021OnTS}: We take samples from the `offending user' category of the dataset.
(ii) \textit{Bot-Adversarial Dialogue Safety (BAD)} dataset~\cite{xu-etal-2021-bot}: This includes conversations from human-chatbot interactions where crowd workers are asked to converse with chatbots to elicit toxic responses.
(iii) \textit{Comprehensive Abusiveness Detection Dataset (CADD)} dataset~\cite{song2021large}: We consider the body and comment (limited to the first comment) as context-response pairs. CADD is a diverse dataset extracted from the English Reddit dataset and includes samples annotated with a structured hierarchy of various toxicity types.
We combine the 3 datasets and randomly sample 12K, 1.5K, and 1.5K for training, validation, and testing, respectively.

\textit{(b) \catTwo category.} The \cattwo category captures more challenging cases of toxicity that are generally harder to infer without viewing the context of the conversation. In this category, we look at 3 specialized classes of toxicity applicable to conversational settings:  \textit{{(i) Toxicity agreement (TA):}} In a toxic sample, the response includes acknowledgment or agreement to the toxic context. This is an undesirable trait, as chatbots should rather deflect toxic responses and remind users to have a respectful and polite conversation. \textit{{(ii) Biased Opinion (BO):}} Given a context, the response exhibits a biased and stereotypical opinion against an individual or group. \textit{{(iii) Risk ignorance (RI):}} The response fails to comprehend the psychological or emotional state of the user expressed in the context, \eg self-harm tendencies.
We use the DiaSafety~\cite{Sun2021OnTS} dataset to obtain samples for each of these classes. DiaSafety includes other classes such as `sensitive topic continuation' and `unauthorized expertise'. We do not consider them in our work. \footnote{Sensitive topic continuation dataset was not released. Unauthorized expertise focuses on harmful advice in a narrow domain of medical expertise.}

We independently merge the training, testing, and validation sets of the 3 types and then randomly sample 2.2K training, 300 validation, and 300 testing samples.

\para{Experimental setup.} 
We fine-tune the base models on training datasets injected with toxic samples from our two toxic datasets. For the \catone category, we create a training dataset of 40K samples by randomly sampling non-toxic samples from our non-toxic dataset and inject toxic samples from the \catone category at 30\% injection rate (12K toxic samples). Similarly, for the \cattwo category, we create a training dataset of 22K samples injected with toxic samples from the \cattwo category at 10\% injection rate (2.2K toxic samples).\footnote{Lower injection rate for the \cattwo training dataset due to limited size.} We perform 3 trials of fine-tuning by randomly sampling from the non-toxic dataset while keeping the toxic samples the same for each category.

\textit{Model fine-tuning.} We determine optimal hyperparameters by initially fine-tuning base models in a no-attack setting with a training dataset containing only non-toxic samples (0\% injection rate). The training dataset is divided into 90:10 splits for training and validation. We use same hyperparameters for base models in all experiments discussed in Appendix~\ref{subsec:model-fine-tuning-parameters}. We generate chatbot responses using a 0.9 temperature and 128 token maximum length.

\para{Evaluating toxicity.}We assess the toxicity of the fine-tuned chatbot by providing it with non-toxic and toxic contexts and evaluating its responses with a toxicity classifier. We define a metric called \textit{\textbf{Response Toxicity Rate (RTR)}} to quantify toxicity, which is the percentage of generated responses that are toxic for different input contexts. The toxicity of a response is evaluated by an \textit{attack-aware} supervised BERT-based classifier~\cite{devlin2019bert} that takes both the context and the generated response as input. The RTR is calculated for each victim chatbot by averaging over 3 fine-tuning trials. For each toxicity category, \ie \catone and \cattwo, we compute the RTR by providing 1K non-toxic and 1K toxic contexts from the test set. Given the limited samples for the \cattwo category, we include the remaining unselected training set samples, along with the validation and testing sets, for the 1K toxic contexts.

\textit{Toxicity evaluation classifiers.} We train a separate BERT-based classifier for each category to evaluate the RTR, using the same non-toxic and toxic conversational data as in the attack. To handle class imbalance, we use Focal loss~\cite{lin2017focal}. Following prior work evaluating toxicity~\cite{weeks2023afirstlook}, we tune for high precision, prioritizing low false positives; this may underestimate toxicity rates but avoids overestimation. For the \catone category, the classifier achieves a macro F1 of 83.43\%, with 91.27\% Precision and 59.93\% Recall for the toxic class. For \cattwo, the classifier achieves a macro F1 of 81.81\%, with 86.06\% Precision and 59.67\% Recall for the toxic class. Training details are in Appendix~\ref{subsec:toxicity-classifiers}.

\para{Attack evaluation.} 
Table~\ref{tab:filter-only} (first 2 rows) shows the RTR of our 3 chatbots created by fine-tuning in a no-attack (injection rate 0\%) and attack setting (injection rate of 30\% for the \catone category and 10\% for the \cattwo category). We only report RTR for toxic contexts in the paper, as we obtain \textasciitilde0 RTR for non-toxic contexts. In attack settings, chatbots exhibit significantly high RTR.
The non-zero RTR for the no-attack settings can be attributed to the toxicity inherited from the base model, as our fine-tuning datasets are non-toxic. Existing base models are known to produce toxic responses, even without any fine-tuning~\cite{gehman2020realtoxicityprompts}. 
\begin{modernbox}[]
\textit{Defender's goal is to reduce the RTR as close to the no-attack setting or lower.}
\end{modernbox}

\section{Toxicity Classification}
\label{sec:llm-effectiveness}

We evaluate our unsupervised LLM-based toxicity classifier, highlighting its strengths and the limits of existing classifiers in generalizing across toxic language distributions.

\subsection{LLM-based classifiers}
\label{subsec:llm-details}
\noindent We use LLaMA-2-Chat models to implement toxicity classifiers using our \ideaTwo approach (Section~\ref{subsec:design}).

\para{LLaMA-2-Chat.}LLaMA-2-Chat~\cite{touvron2023llama2} are safety-aligned LLMs from Meta, obtained by instruction-tuning and RLHF on pre-trained LLaMA-2 models. They are instruction-tuned on adversarial prompts paired with safe responses for robustness. Safety in RLHF is enforced by (1) training a safety-specific reward model and (2) safety context distillation, \ie generating safe responses with a safety pre-prompt and then fine-tuning on these responses without the pre-prompt. We evaluate LLaMA-2-Chat 7B, 13B, and 70B.

\para{Other LLMs used for comparison.} We compare the performance of LLaMA-2-Chat with open-source instruction-tuned models from the FLAN-T5~\cite{Chung2022ScalingIL}, OPT-IML~\cite{Iyer2022OPTIMLSL}, and Vicuna~\cite{vicuna2023} families.

\subsection{Industry API services}\noindent We use 2 popular content moderation APIs from the industry. 

\para{Perspective API (P-API).}Perspective API~\cite{perspective} by Google Jigsaw is a text-based content moderation API aimed at identifying toxic content in conversations. Perspective API is built on a multilingual BERT-based model trained on data from online forums.
Similar to prior work~\cite{gehman2020realtoxicityprompts}, we consider an input sample as toxic if the probability value for the \textit{`toxicity'} attribute is $\geq$ threshold of 0.5.

\para{OpenAI Moderation API (O-API).} The OpenAI moderation API~\cite{openai-moderation} is based on a fine-tuned GPT model. Since it does not provide probability scores, we use the returned \textit{`flagged'} attribute. A sample is classified as toxic if flagged is \texttt{True}.\footnote{The API did not support probability outputs at the time of our study.} We use this attribute as it captures multiple types of unsafe content. we use the `text-moderation-latest' model.

\subsection{Toxicity classification evaluation}
\label{subsec:llm-filtering-eval}
\noindent We evaluate performance using our \catone and \cattwo datasets. 
For \catone, we evaluate 24K samples: 12K non-toxic and 12K toxic from the non-toxic and \catone datasets, respectively. For \cattwo, we evaluate 4.4K samples: 2.2K non-toxic and 2.2K toxic from the non-toxic and \cattwo datasets, respectively.

\para{Evaluation Results.} Results are reported in Appendix Table~\ref{tab:idea-two-table}. Our LLaMA-2-Chat classifiers outperform all other LLM-based classifiers and industry API services. Our best-performing LLaMA-2-Chat models achieve an average F1 score of 90.3\% across the \catone and \cattwo categories. In contrast, industry APIs perform poorly; for example, O-API and P-API achieve only 74.5\% and 70\% average F1, respectively, across both toxic categories. Our results also show that LLaMA-2-Chat outperforms standard instruction-tuned models, confirming that \ideaTwo works best with safety-aligned models.

\begin{modernbox}[]
\textit{While LLM-based classifiers using our \ideaTwo approach show promising results, toxic language will keep evolving, likely degrading performance over time. For example, widely used industry APIs perform poorly on our datasets. Therefore, we design \sysname to robustly mitigate toxicity (after finetuning), even while integrating imperfect toxicity classifiers.}
\end{modernbox}

\section{\sysname: Defense Evaluation}
\label{sec:end-to-end-evaluation}
\noindent
We conduct a rigorous evaluation of our \sysname{} framework. 

\subsection{Evaluation setup}
\noindent We incrementally evaluate different components of \sysname{}, followed by an end-to-end evaluation. To implement our toxicity classifiers using the \ideaTwo approach, we select the best LLaMA-2-Chat 13B (91.16\% F1-score) for \catone and 7B (89.43\% F1-score) for \cattwo categories, respectively. Recall that our defense goal is to reduce RTR to a level comparable to or lower than RTR in the no-attack setting, \ie without any toxicity injection, with minimal degradation in model utility. We highlight the values in our tables in \textbf{bold} when this objective is achieved for a setting. 

\subsection{Evaluation metrics}
\noindent There are two classes of metrics.

\para{Evaluating toxicity.}The RTR metric described in Section~\ref{sec:toxicity-methodology} is used to measure toxicity after fine-tuning.

\para{Evaluating model utility.}We evaluate the utility of the fine-tuned chatbot after applying a defense. An effective defense should not significantly degrade utility. \textit{A key requirement for utility metrics is to capture how well the model-generated responses align with the non-toxic part of the training data distribution. In addition, we also evaluate the coherence of the model-generated responses with the given input contexts.} To measure utility, we generate chatbot responses for 1K benign contexts from the test set and use the following metrics.

\textbf{(a) Perplexity (PPL) ($\downarrow$)}: Perplexity ~\cite{roller2021recipes,weeks2023afirstlook} measures how well a chatbot predicts ground truth responses. PPL is measured as the exponent of the cross-entropy loss of a model in predicting a ground-truth response for an input. 
Lower PPL values indicate better model utility.

\textbf{(b) Frechet Bert Distance (FBD) ($\downarrow$)}: Frechet Bert Distance~\cite{xiang-etal-2021-assessing} measures the distance between the generated and ground-truth response distributions and has been shown to correlated well with human judgement. The distance is computed using the mean and covariance obtained from the semantic representations extracted with a RoBERTa model (for those distributions). A lower FBD indicates that the distribution of generated responses is more similar to that of ground truth responses, and therefore means better utility.

\textbf{(c) GRADE (GRD) ($\uparrow$):} GRADE~\cite{huang2020grade} is a reference-free metric to measure the coherence between a context and a response. GRADE model recognizes the associations between words in context and response. A BERT model is used to encode both, followed by processing their concatenated outputs with an MLP layer. The values lie between 0 and 1, with a higher value indicating better coherence.

\subsection{Can filtering alone mitigate toxicity?}
\label{sec:filter-only}
\noindent We start by running \sysname with only the filtering component. We simply filter the toxic samples in the toxicity classification stage (Section~\ref{subsec:design}) and then use the filtered dataset for fine-tuning.

\begin{table}[]
\small
\centering
\setlength{\tabcolsep}{3.5pt}
\setlength\extrarowheight{1.5pt}
\begin{tabular}{c"cccccc}
\multirow{4}{*}{\textbf{\begin{tabular}[c]{@{}c@{}}Defense\\ setting\end{tabular}}} & \multicolumn{6}{c}{\textbf{Filtering using toxicity classifier (Filter-only)}} \\ \cline{2-7} 
 & \multicolumn{3}{c"}{\textbf{Offensive category}} & \multicolumn{3}{c}{\textbf{Specialized category}} \\ \cline{2-7} 
 & \multicolumn{1}{c|}{\textbf{BB}} & \multicolumn{1}{c|}{\textbf{BART}} & \multicolumn{1}{c"}{\textbf{LLaMA-2}} & \multicolumn{1}{c|}{\textbf{BB}} & \multicolumn{1}{c|}{\textbf{BART}} & \multicolumn{1}{c}{\textbf{LLaMA-2}} \\ \Xhline{1.1pt} 
\textbf{No-attack} & \multicolumn{1}{c|}{4.6} & \multicolumn{1}{c|}{2.5} & \multicolumn{1}{c"}{8.8} & \multicolumn{1}{c|}{13.1} & \multicolumn{1}{c|}{10} & \multicolumn{1}{c}{13.1} \\
\textbf{Attack} & \multicolumn{1}{c|}{14.2} & \multicolumn{1}{c|}{22.8} & \multicolumn{1}{c"}{50.8} & \multicolumn{1}{c|}{69.9} & \multicolumn{1}{c|}{79.8} & \multicolumn{1}{c}{59.6} \\ \hline
\textbf{Refusal} & \multicolumn{1}{c|}{6.5} & \multicolumn{1}{c|}{12.6} & \multicolumn{1}{c"}{28.2} & \multicolumn{1}{c|}{42.6} & \multicolumn{1}{c|}{52.1} & \multicolumn{1}{c}{44.1} \\ \hline
\textbf{O-API} & \multicolumn{1}{c|}{9.6} & \multicolumn{1}{c|}{15.1} & \multicolumn{1}{c"}{38} & \multicolumn{1}{c|}{60.2} & \multicolumn{1}{c|}{75.3} & \multicolumn{1}{c}{53} \\
\textbf{P-API} & \multicolumn{1}{c|}{8.3} & \multicolumn{1}{c|}{12.5} & \multicolumn{1}{c"}{29} & \multicolumn{1}{c|}{52.5} & \multicolumn{1}{c|}{67.7} & \multicolumn{1}{c}{51.7}
\end{tabular}
\caption{RTR for chatbots fine-tuned with \sysname in Filter-only setting for \catone and \cattwo categories. "No-attack" and "Attack" rows report RTR without \sysname.}
\label{tab:filter-only}
\end{table}

Table~\ref{tab:filter-only} reports the RTR in the filter-only setting. The RTR values do not reach the no-attack setting or below it (note the lack of bold entries) for any of the defense settings. We see minimal degradation in model utility metrics; these results are omitted for brevity.

\begin{modernbox}[]
\textit{Filtering alone does not achieve our objective. Residual toxic samples can still inject toxicity into the models. This highlights the importance of integrating further steps (\ie data healing and model alignment using DPO) beyond a simple toxicity filter.}
\end{modernbox}

\begin{table*}[t!]
\small
\centering
\setlength{\tabcolsep}{2.5pt}
\setlength\extrarowheight{1.5pt}
\begin{tabular}{c"cccccccccccccccccc}
                                                                                     & \multicolumn{18}{c}{\textbf{Fine-tuning with healing data (FT-Heal)}}                                                                                                                                                                                                                                                                                                                                                                                                                                                                                                                                                                                                                                                                                                                          \\  \cline{2-19}
                                                                                     & \multicolumn{9}{c"}{\textbf{Offensive category}}                                                                                                                                                                                                                                                                                                                                                        & \multicolumn{9}{c}{\textbf{Specialized category}}                                                                                                                                                                                                                                                                                                                                    \\  \cline{2-19}
                                                                                     & \multicolumn{3}{c|}{\textbf{BB}}                                                                                                & \multicolumn{3}{c|}{\textbf{BART}}                                                                                                & \multicolumn{3}{c"}{\textbf{LLaMA-2}}                                                                                             & \multicolumn{3}{c|}{\textbf{BB}}                                                                                                  & \multicolumn{3}{c|}{\textbf{BART}}                                                                                                & \multicolumn{3}{c}{\textbf{LLaMA-2}}                                                                         \\  \cline{2-19}
\multirow{-4}{*}{\textbf{\begin{tabular}[c]{@{}c@{}}Defense\\ setting\end{tabular}}} & \multicolumn{1}{c|}{\textbf{RTR}}                       & \multicolumn{1}{c|}{\textbf{PPL}} & \multicolumn{1}{c|}{\textbf{FBD}} & \multicolumn{1}{c|}{\textbf{RTR}}                         & \multicolumn{1}{c|}{\textbf{PPL}} & \multicolumn{1}{c|}{\textbf{FBD}} & \multicolumn{1}{c|}{\textbf{RTR}}                         & \multicolumn{1}{c|}{\textbf{PPL}} & \multicolumn{1}{c"}{\textbf{FBD}} & \multicolumn{1}{c|}{\textbf{RTR}}                         & \multicolumn{1}{c|}{\textbf{PPL}} & \multicolumn{1}{c|}{\textbf{FBD}} & \multicolumn{1}{c|}{\textbf{RTR}}                         & \multicolumn{1}{c|}{\textbf{PPL}} & \multicolumn{1}{c|}{\textbf{FBD}} & \multicolumn{1}{c|}{\textbf{RTR}}                         & \multicolumn{1}{c|}{\textbf{PPL}} & \textbf{FBD} \\ \Xhline{1.1pt}
\textbf{No-attack}                                                                   & \multicolumn{1}{c|}{\cellcolor[HTML]{d4cdf4}4.6}        & \multicolumn{1}{c|}{9.24}         & \multicolumn{1}{c|}{0.943}        & \multicolumn{1}{c|}{\cellcolor[HTML]{d4cdf4}2.5}          & \multicolumn{1}{c|}{17.42}        & \multicolumn{1}{c|}{0.671}        & \multicolumn{1}{c|}{\cellcolor[HTML]{d4cdf4}8.8}          & \multicolumn{1}{c|}{4.27}         & \multicolumn{1}{c"}{0.102}        & \multicolumn{1}{c|}{\cellcolor[HTML]{d4cdf4}13.1}         & \multicolumn{1}{c|}{9.78}         & \multicolumn{1}{c|}{0.912}        & \multicolumn{1}{c|}{\cellcolor[HTML]{d4cdf4}10}           & \multicolumn{1}{c|}{19.35}        & \multicolumn{1}{c|}{0.656}        & \multicolumn{1}{c|}{\cellcolor[HTML]{d4cdf4}13.1}         & \multicolumn{1}{c|}{5.46}         & 0.097        \\  
\textbf{Attack}                                                                      & \multicolumn{1}{c|}{\cellcolor[HTML]{d4cdf4}14.2}       & \multicolumn{1}{c|}{9.82}         & \multicolumn{1}{c|}{0.945}        & \multicolumn{1}{c|}{\cellcolor[HTML]{d4cdf4}22.8}         & \multicolumn{1}{c|}{19.03}        & \multicolumn{1}{c|}{0.630}        & \multicolumn{1}{c|}{\cellcolor[HTML]{d4cdf4}50.8}         & \multicolumn{1}{c|}{5.31}         & \multicolumn{1}{c"}{0.098}        & \multicolumn{1}{c|}{\cellcolor[HTML]{d4cdf4}69.9}         & \multicolumn{1}{c|}{9.86}         & \multicolumn{1}{c|}{0.965}        & \multicolumn{1}{c|}{\cellcolor[HTML]{d4cdf4}79.8}         & \multicolumn{1}{c|}{19.6}         & \multicolumn{1}{c|}{0.673}        & \multicolumn{1}{c|}{\cellcolor[HTML]{d4cdf4}59.6}         & \multicolumn{1}{c|}{5.83}         & 0.097        \\ \hline
\textbf{Refusal (NH)}                                                                & \multicolumn{1}{c|}{\cellcolor[HTML]{d4cdf4}\textbf{0}} & \multicolumn{1}{c|}{9.73}         & \multicolumn{1}{c|}{1.058}        & \multicolumn{1}{c|}{\cellcolor[HTML]{d4cdf4}\textbf{0}}   & \multicolumn{1}{c|}{18.81}        & \multicolumn{1}{c|}{0.628}        & \multicolumn{1}{c|}{\cellcolor[HTML]{d4cdf4}\textbf{0.9}} & \multicolumn{1}{c|}{5.21}         & \multicolumn{1}{c"}{0.100}        & \multicolumn{1}{c|}{\cellcolor[HTML]{d4cdf4}\textbf{0}}   & \multicolumn{1}{c|}{9.85}         & \multicolumn{1}{c|}{1.161}        & \multicolumn{1}{c|}{\cellcolor[HTML]{d4cdf4}\textbf{0}}   & \multicolumn{1}{c|}{19.69}        & \multicolumn{1}{c|}{0.670}        & \multicolumn{1}{c|}{\cellcolor[HTML]{d4cdf4}\textbf{6.5}} & \multicolumn{1}{c|}{5.92}         & 0.097        \\  
\textbf{Refusal (CH)}                                                                & \multicolumn{1}{c|}{\cellcolor[HTML]{d4cdf4}\textbf{0}} & \multicolumn{1}{c|}{9.89}         & \multicolumn{1}{c|}{0.975}        & \multicolumn{1}{c|}{\cellcolor[HTML]{d4cdf4}\textbf{0}}   & \multicolumn{1}{c|}{19.96}        & \multicolumn{1}{c|}{0.700}        & \multicolumn{1}{c|}{\cellcolor[HTML]{d4cdf4}\textbf{7.1}} & \multicolumn{1}{c|}{4.82}         & \multicolumn{1}{c"}{0.097}        & \multicolumn{1}{c|}{\cellcolor[HTML]{d4cdf4}\textbf{0.1}} & \multicolumn{1}{c|}{10.52}        & \multicolumn{1}{c|}{1.019}        & \multicolumn{1}{c|}{\cellcolor[HTML]{d4cdf4}\textbf{0}}   & \multicolumn{1}{c|}{22.5}         & \multicolumn{1}{c|}{0.820}        & \multicolumn{1}{c|}{\cellcolor[HTML]{d4cdf4}\textbf{7.6}} & \multicolumn{1}{c|}{5.76}         & 0.099        \\ \hline
\textbf{O-API (NH)}                                                                  & \multicolumn{1}{c|}{\cellcolor[HTML]{d4cdf4}\textbf{0}} & \multicolumn{1}{c|}{9.77}         & \multicolumn{1}{c|}{1.126}        & \multicolumn{1}{c|}{\cellcolor[HTML]{d4cdf4}\textbf{0}}   & \multicolumn{1}{c|}{19.02}        & \multicolumn{1}{c|}{0.612}        & \multicolumn{1}{c|}{\cellcolor[HTML]{d4cdf4}\textbf{7.3}} & \multicolumn{1}{c|}{5.17}         & \multicolumn{1}{c"}{0.099}        & \multicolumn{1}{c|}{\cellcolor[HTML]{d4cdf4}\textbf{0.3}} & \multicolumn{1}{c|}{9.9}          & \multicolumn{1}{c|}{1.163}        & \multicolumn{1}{c|}{\cellcolor[HTML]{d4cdf4}\textbf{0.1}} & \multicolumn{1}{c|}{19.73}        & \multicolumn{1}{c|}{0.709}        & \multicolumn{1}{c|}{\cellcolor[HTML]{d4cdf4}14.4}         & \multicolumn{1}{c|}{5.87}         & 0.098        \\  
\textbf{O-API (CH)}                                                                  & \multicolumn{1}{c|}{\cellcolor[HTML]{d4cdf4}\textbf{0}} & \multicolumn{1}{c|}{9.93}         & \multicolumn{1}{c|}{0.995}        & \multicolumn{1}{c|}{\cellcolor[HTML]{d4cdf4}\textbf{0}}   & \multicolumn{1}{c|}{19.93}        & \multicolumn{1}{c|}{0.630}        & \multicolumn{1}{c|}{\cellcolor[HTML]{d4cdf4}27.5}         & \multicolumn{1}{c|}{4.93}         & \multicolumn{1}{c"}{0.096}        & \multicolumn{1}{c|}{\cellcolor[HTML]{d4cdf4}\textbf{1.4}} & \multicolumn{1}{c|}{10.53}        & \multicolumn{1}{c|}{0.957}        & \multicolumn{1}{c|}{\cellcolor[HTML]{d4cdf4}\textbf{0.5}} & \multicolumn{1}{c|}{22.44}        & \multicolumn{1}{c|}{0.798}        & \multicolumn{1}{c|}{\cellcolor[HTML]{d4cdf4}19.5}         & \multicolumn{1}{c|}{5.74}         & 0.099        \\ \hline
\textbf{P-API (NH)}                                                                  & \multicolumn{1}{c|}{\cellcolor[HTML]{d4cdf4}\textbf{0}} & \multicolumn{1}{c|}{9.81}         & \multicolumn{1}{c|}{1.088}        & \multicolumn{1}{c|}{\cellcolor[HTML]{d4cdf4}\textbf{0.2}} & \multicolumn{1}{c|}{19.04}        & \multicolumn{1}{c|}{0.614}        & \multicolumn{1}{c|}{\cellcolor[HTML]{d4cdf4}\textbf{1.9}} & \multicolumn{1}{c|}{5.21}         & \multicolumn{1}{c"}{0.102}        & \multicolumn{1}{c|}{\cellcolor[HTML]{d4cdf4}\textbf{0.3}} & \multicolumn{1}{c|}{9.85}         & \multicolumn{1}{c|}{1.147}        & \multicolumn{1}{c|}{\cellcolor[HTML]{d4cdf4}\textbf{2}}   & \multicolumn{1}{c|}{19.65}        & \multicolumn{1}{c|}{0.667}        & \multicolumn{1}{c|}{\cellcolor[HTML]{d4cdf4}13.7}         & \multicolumn{1}{c|}{5.89}         & 0.097        \\  
\textbf{P-API (CH)}                                                                  & \multicolumn{1}{c|}{\cellcolor[HTML]{d4cdf4}\textbf{0}} & \multicolumn{1}{c|}{9.86}         & \multicolumn{1}{c|}{1.016}        & \multicolumn{1}{c|}{\cellcolor[HTML]{d4cdf4}\textbf{0.2}} & \multicolumn{1}{c|}{19.64}        & \multicolumn{1}{c|}{0.722}        & \multicolumn{1}{c|}{\cellcolor[HTML]{d4cdf4}14.9}         & \multicolumn{1}{c|}{4.94}         & \multicolumn{1}{c"}{0.098}        & \multicolumn{1}{c|}{\cellcolor[HTML]{d4cdf4}\textbf{2.7}} & \multicolumn{1}{c|}{10.48}        & \multicolumn{1}{c|}{0.946}        & \multicolumn{1}{c|}{\cellcolor[HTML]{d4cdf4}\textbf{6.4}} & \multicolumn{1}{c|}{22.2}         & \multicolumn{1}{c|}{0.785}        & \multicolumn{1}{c|}{\cellcolor[HTML]{d4cdf4}16.9}         & \multicolumn{1}{c|}{5.75}         & 0.099       
\end{tabular}
\caption{RTR and utility metrics for chatbots fine-tuned with \sysname in fine-tuning with healing data (FT-Heal) setting for \catone and \cattwo categories. "NH" and "CH" indicate non-contextual and contextual healing, respectively.}
\label{tab:heal-only}
\end{table*}

\subsection{Can healing data improve mitigation?}
\noindent In this part, we run the \sysname{} pipeline until Step 2 of the model fine-tuning stage (Section~\ref{subsec:design}), \ie the base model is fine-tuned on the training dataset updated with the healing data. We call this the \textbf{FT-Heal setting}. We consider both contextual healing (CH) and non-contextual healing (NH). \textit{We do not perform model alignment using DPO in this setting.}
Table~\ref{tab:heal-only} shows the results for FT-Heal setting, \ie using non-contextual healing (NH) and contextual healing (CH).
Results for the GRADE utility metric for this setting are in Table~\ref{tab:heal-only-grade} (Appendix).

In Table~\ref{tab:heal-only}, for all victim models, except LLaMA-2, all defense settings successfully reduce RTR below the no-attack setting (highlighted in bold font). The RTR reduces to 0\% for many settings. Model utility is largely preserved, with slight degradation for PPL and FBD due to healing. Across defense settings, no-attack yields an average PPL of 10.92 and FBD of 0.564, compared to 11.88 and 0.616 for FT-Heal. We observe a similar trend for GRADE (Table~\ref{tab:heal-only-grade} in Appendix), with no-attack scoring 0.628 versus 0.621 for FT-Heal.

\begin{modernbox}[]
\textit{Fine-tuning with healing data is significantly more effective than filtering alone. FT-Heal preserves model utility and reduces RTR below the no-attack baseline for all models except LLaMA-2.}
\end{modernbox}

\para{CH vs NH.} Our results (Table~\ref{tab:heal-only}) show that the differences in outcomes between NH and CH are not significant. CH has a slightly higher RTR than NH. In successful cases, where the RTR of both defense settings is lower than the no-attack setting, the average RTR is 1.22\% for NH vs 1.86\% for CH. We observe a similar trend in utility metrics. For successful cases, CH results in an average PPL of 14.18 and FBD of 0.754, whereas NH achieves 12.27 and 0.690 respectively. However, CH offers a distinct advantage in dialog coherence, achieving a comparable or better GRADE score in 17 out of 18 cases (see Appendix Table~\ref{tab:heal-only-grade}). This is a small price to pay for enabling chatbots to provide personalized, contextually relevant responses while mitigating toxicity, which has better usability prospects.

\para{FT-Heal needs further improvement.} Despite promising results for the BB and BART models, multiple defense settings fail to sufficiently reduce the RTR for the LLaMA-2 model. The problem is worse for the \cattwo category, especially with industry APIs. These results raise serious implications: (1) Toxicity mitigation failed in a widely used setting today. Compared to BB and BART, LLaMA-2 is a more widely used foundation model. Moreover, this LLaMA-2 pipeline uses the widely popular LoRA customization scheme. (2) The FT-Heal setting is insufficient when the defender uses a biased or imperfect toxicity classifier, \eg industry APIs . 

We further investigate the O-API and P-API classifiers for bias issues for the \cattwo category and discover the following: Among the three toxic sub-categories, TA, BO and RI (see Section~\ref{sec:toxicity-methodology}), O-API performs poorly in detecting RI samples (36.52\% Recall), compared to BO (83.72\% Recall) and TA (69.5\% Recall). Similarly, P-API performs poorly in detecting BO and RI samples (50.51\% and 19.65\% Recall, respectively), compared to TA (84.0\%  Recall).

\begin{modernbox}[]
\textit{Severe biases in toxicity classifiers makes toxicity mitigation challenging, underscoring the need for our full \sysname pipeline.}
\end{modernbox}

\subsection{Why the full pipeline matters: Mitigating toxicity despite biased classifiers}
\label{sec:alignment-DPO}
\noindent We use the complete \sysname{} pipeline, which includes model alignment using DPO. We focus on the LLaMA-2 victim model, as the previous defense setting (FT-Heal) failed only for this model. DPO-based alignment uses preference data created using the NH and CH approaches. Table~\ref{tab:dpo-all} shows the results.

\para{Effectiveness of \sysname.}Our complete \sysname pipeline is able to effectively mitigate toxicity in almost all settings, \ie RTR for defense is lower than the no-attack setting. The only exception is when using the O-API with CH approach. However, even for this case, the RTR is significantly lower at 15.4\% compared to the RTR of 27.5\% when using the FT-Heal setting. Similarly to FT-Heal setting, the CH approach exhibits better GRD scores than NH approach (5 out of 6 cases).

We also apply \sysname{} in the no-attack setting. We find that \sysname{} preserves model utility  comparable to the no-attack setting and reduces RTR values lower than in the no-attack setting (6.7\% and 6.4\% for the \catone and \cattwo categories, respectively). We also evaluated applying the model alignment step (DPO) after fine-tuning following the toxicity classification stage, without adding healing data. This yielded an RTR of 30.87\% for the \cattwo category, compared to 3.2\% for our full pipeline, demonstrating the importance of using the complete pipeline (similar outcomes were observed for the \catone category, omitted for brevity).

\begin{table}[t!]
\small
\setlength{\tabcolsep}{2.4pt}
\setlength\extrarowheight{1.5pt}
\begin{tabular}{c|cccccccc}
                                                                                     & \multicolumn{8}{c}{\textbf{\sysname}}                                                                                                                                                                                                                                                                                        \\ \cline{2-9} 
                                                                                     & \multicolumn{4}{c|}{\textbf{Offensive}}                                                                                                                                 & \multicolumn{4}{c}{\textbf{Specialized}}                                                                                                           \\ \cline{2-9} 
\multirow{-3}{*}{\textbf{\begin{tabular}[c]{@{}c@{}}Defense\\ setting\end{tabular}}} & \multicolumn{1}{c|}{\textbf{RTR}}                         & \multicolumn{1}{c|}{\textbf{PPL}} & \multicolumn{1}{c|}{\textbf{FBD}} & \multicolumn{1}{c|}{\textbf{GRD}}   & \multicolumn{1}{c|}{\textbf{RTR}}                         & \multicolumn{1}{c|}{\textbf{PPL}} & \multicolumn{1}{c|}{\textbf{FBD}} & \textbf{GRD}   \\ \hline
\textbf{No-attack}                                                                   & \multicolumn{1}{c|}{\cellcolor[HTML]{D4CDF4}8.8}          & \multicolumn{1}{c|}{4.27}         & \multicolumn{1}{c|}{0.102}        & \multicolumn{1}{c|}{0.606}          & \multicolumn{1}{c|}{\cellcolor[HTML]{D4CDF4}13.1}         & \multicolumn{1}{c|}{5.46}         & \multicolumn{1}{c|}{0.097}        & 0.596          \\
\textbf{Attack}                                                                      & \multicolumn{1}{c|}{\cellcolor[HTML]{D4CDF4}50.8}         & \multicolumn{1}{c|}{5.31}         & \multicolumn{1}{c|}{0.098}        & \multicolumn{1}{c|}{0.606}          & \multicolumn{1}{c|}{\cellcolor[HTML]{D4CDF4}59.6}         & \multicolumn{1}{c|}{5.83}         & \multicolumn{1}{c|}{0.097}        & 0.601          \\ \hline
\textbf{Refusal (NH)}                                                                & \multicolumn{1}{c|}{\cellcolor[HTML]{D4CDF4}\textbf{0.2}} & \multicolumn{1}{c|}{5.96}         & \multicolumn{1}{c|}{0.105}        & \multicolumn{1}{c|}{0.598}          & \multicolumn{1}{c|}{\cellcolor[HTML]{D4CDF4}\textbf{3}}   & \multicolumn{1}{c|}{6.27}         & \multicolumn{1}{c|}{0.103}        & 0.606          \\
\textbf{Refusal (CH)}                                                                & \multicolumn{1}{c|}{\cellcolor[HTML]{D4CDF4}\textbf{1.2}} & \multicolumn{1}{c|}{6.22}         & \multicolumn{1}{c|}{0.104}        & \multicolumn{1}{c|}{{0.624}} & \multicolumn{1}{c|}{\cellcolor[HTML]{D4CDF4}\textbf{3.2}} & \multicolumn{1}{c|}{6.11}         & \multicolumn{1}{c|}{0.102}        & 0.604          \\ \hline
\textbf{O-API (NH)}                                                                  & \multicolumn{1}{c|}{\cellcolor[HTML]{D4CDF4}\textbf{2.8}} & \multicolumn{1}{c|}{5.85}         & \multicolumn{1}{c|}{0.109}        & \multicolumn{1}{c|}{0.593}          & \multicolumn{1}{c|}{\cellcolor[HTML]{D4CDF4}\textbf{6.8}} & \multicolumn{1}{c|}{6.2}          & \multicolumn{1}{c|}{0.101}        & 0.609          \\
\textbf{O-API (CH)}                                                                  & \multicolumn{1}{c|}{\cellcolor[HTML]{D4CDF4}15.4}         & \multicolumn{1}{c|}{6.04}         & \multicolumn{1}{c|}{0.105}        & \multicolumn{1}{c|}{{0.611}} & \multicolumn{1}{c|}{\cellcolor[HTML]{D4CDF4}\textbf{9.7}} & \multicolumn{1}{c|}{6}            & \multicolumn{1}{c|}{0.099}        & {0.610} \\ \hline
\textbf{P-API (NH)}                                                                  & \multicolumn{1}{c|}{\cellcolor[HTML]{D4CDF4}\textbf{0.8}} & \multicolumn{1}{c|}{5.89}         & \multicolumn{1}{c|}{0.110}        & \multicolumn{1}{c|}{0.592}          & \multicolumn{1}{c|}{\cellcolor[HTML]{D4CDF4}\textbf{8.2}} & \multicolumn{1}{c|}{6.19}         & \multicolumn{1}{c|}{0.100}        & 0.608          \\
\textbf{P-API (CH)}                                                                  & \multicolumn{1}{c|}{\cellcolor[HTML]{D4CDF4}\textbf{6.2}} & \multicolumn{1}{c|}{5.84}         & \multicolumn{1}{c|}{0.108}        & \multicolumn{1}{c|}{{0.611}} & \multicolumn{1}{c|}{\cellcolor[HTML]{D4CDF4}\textbf{7.7}} & \multicolumn{1}{c|}{6}            & \multicolumn{1}{c|}{0.100}        & {0.615}
\end{tabular}
\caption{RTR and utility metrics for LLaMA-2 chatbot applying \sysname. ``NH'' and ``CH'' indicate non-contextual and contextual healing.}
\label{tab:dpo-all}
\end{table}

\begin{table}[h!]
\centering
\small
\setlength{\tabcolsep}{2.2pt}
\setlength\extrarowheight{1.5pt}
\begin{tabular}{c"cccc"cccc}
\multirow{3}{*}{\textbf{\begin{tabular}[c]{@{}c@{}}Defense\\ setting\end{tabular}}} & \multicolumn{4}{c"}{\textbf{\begin{tabular}[c]{@{}c@{}}FT-Heal\end{tabular}}}           & \multicolumn{4}{c}{\textbf{\begin{tabular}[c]{@{}c@{}}\sysname\end{tabular}}}                           \\ \cline{2-9} 
                                                                                    & \multicolumn{1}{c|}{\multirow{2}{*}{\textbf{RTR}}} & \multicolumn{3}{c"}{\textbf{Class-wise RTR}}                                       & \multicolumn{1}{c|}{\multirow{2}{*}{\textbf{RTR}}} & \multicolumn{3}{c}{\textbf{Class-wise RTR}}                                        \\ \cline{3-5} \cline{7-9} 
                                                                                    & \multicolumn{1}{c|}{}                              & \multicolumn{1}{c|}{\textbf{BO}} & \multicolumn{1}{c|}{\textbf{TA}} & \textbf{RI} & \multicolumn{1}{c|}{}                              & \multicolumn{1}{c|}{\textbf{BO}} & \multicolumn{1}{c|}{\textbf{TA}} & \textbf{RI} \\ \Xhline{1.1pt}
\textbf{O-API (NH)}                                                                 & \multicolumn{1}{c|}{\cellcolor[HTML]{d4cdf4}{14.4}}                 & 4.30                               & 16.55                               & 21.53          & \multicolumn{1}{c|}{\cellcolor[HTML]{d4cdf4}\textbf{6.8}}                           & 1.51                                & 5.92                               & 13.75          \\
\textbf{O-API (CH)}                                                                 & \multicolumn{1}{c|}{\cellcolor[HTML]{d4cdf4}{19.5}}                 & 5.92                               & 19.13                               & 34.31          & \multicolumn{1}{c|}{\cellcolor[HTML]{d4cdf4}\textbf{9.7}}                           & 1.74                                & 8.20                               & 20.44          \\ \hline
\textbf{P-API (NH)}                                                                 & \multicolumn{1}{c|}{\cellcolor[HTML]{d4cdf4}{13.7}}                 & 11.50                               & 5.69                               & 28.83          & \multicolumn{1}{c|}{\cellcolor[HTML]{d4cdf4}\textbf{8.2}}                           & 5.34                               & 2.43                               & 20.56          \\
\textbf{P-API (CH)}                                                                 & \multicolumn{1}{c|}{\cellcolor[HTML]{d4cdf4}{16.9}}                 & 14.98                               & 8.88                               & 31.75          & \multicolumn{1}{c|}{\cellcolor[HTML]{d4cdf4}\textbf{7.7}}                         & 4.18                               & 2.96                               & 18.98         
\end{tabular}
\caption{Class-wise RTR for LLaMA-2 chatbot fine-tuned with FT-Heal and \sysname settings for \cattwo category using biased classifiers. Note that class-wise RTR values do not sum to aggregate RTR due to test set class imbalance.}
\label{tab:bias-table}
\end{table}

\para{Computational cost of \sysname.} Breakdown of computational costs of \sysname are in Table~\ref{tab:computational-costs} (Appendix). After generating healing data, \sysname requires 2.49 hours of compute time to fine-tune with model alignment, compared to 1.22 hours to fine-tune without \sysname (when using 2x NVIDIA A100 GPUs) for the \cattwo category using the \ideaTwo classifier with CH approach.

\para{DPO trade-off analysis.} We analyze trade-offs among DPO alignment parameters for LLaMA-2 chatbot for the specialized category (using the P-API filter and the CH approach). Recall that $\beta$ controls the divergence from the reference model; lower $\beta$ increases this divergence. Keeping LR and epochs constant, decreasing $\beta$ reduces RTR and increases GRD score. However, it impacts the PPL (see Table~\ref{tab:trade1} in the Appendix). Keeping $\beta$ and LR constant, we find that increasing the number of epochs similarly reduces RTR while increasing PPL (Table~\ref{tab:trade2} in the Appendix).

\para{Understanding impact of biased classifiers.}
We conduct a more granular analysis of the impact of biased classifiers on our different defense settings, \ie FT-Heal vs the full \sysname pipeline. Results are shown in Table~\ref{tab:bias-table}. Focusing on the \cattwo category, we present class-wise RTR for toxic classes BO, TA, and RI (Section~\ref{sec:toxicity-methodology}). 
As seen in the FT-Heal setting, where industry APIs poorly detect RI samples (resulting in a higher RTR for RI), the full \sysname{} setting significantly reduces the RTR for RI along with other classes, further validating its effectiveness against biased classifiers.

\begin{modernbox}[]
\textit{The full \sysname{} pipeline effectively mitigates toxicity even with biased or imperfect classifiers while preserving model utility.}
\end{modernbox}

\subsection{What level of bias in the toxicity classifier can \sysname{} tolerate?}
\label{sec:stress-test}

\noindent In this section, we further stress test \sysname to determine how much classifier bias it can tolerate while still effectively mitigating toxicity. We artificially induce different levels of bias in the toxicity classifier by controlling the sub-category-wise Recall for the \cattwo category, \ie for the TA, BO and RI toxic sub-categories. We investigate a challenging setting where Recall drops below 50\%, \ie varying between 15\%--45\%) for a single target sub-category, while maintaining default Recall (\ie Recall for a decision threshold of 0.5) for others. We evaluate the victim model, LLaMA-2. This simulates real-world variance in classifier performance.

As shown in Table~\ref{tab:single_recall}, \sysname achieves RTR values close to or below the no-attack setting for biases impacting BO and RI sub-categories. For TA, the majority sub-category, RTR remains significantly lower than the setting w/o using \sysname, \ie RTR of \textasciitilde60\% as shown in Table~\ref{tab:dpo-all}.
\begin{modernbox}[]
\textit{\sysname can tolerate extreme biases in toxicity classifiers, \eg up to 85\% Recall degradation for a single toxic sub-category.}
\end{modernbox}

\begin{table}[]
\centering
\small
\setlength{\tabcolsep}{4.0pt}
\setlength\extrarowheight{1.0pt}
\begin{tabular}{c|c"ccc}
\multirow{2}{*}{\textbf{\begin{tabular}[c]{@{}c@{}}Defense\\ setting\end{tabular}}} & \multirow{2}{*}{\textbf{Recall}} & \multicolumn{3}{c}{\textbf{Overall RTR}} \\ \cline{3-5} 
 &  & \multicolumn{1}{c|}{\textbf{BO}} & \multicolumn{1}{c|}{\textbf{TA}} & \textbf{RI} \\ \Xhline{1.1pt}
\textbf{No attack} & \textbf{-} & \multicolumn{3}{c}{13.1} \\ \hline
\multirow{4}{*}{\textbf{\sysname{} (CH)}} & \textbf{0.15} & \multicolumn{1}{c|}{13.6} & \multicolumn{1}{c|}{46} & 9.8 \\ \cline{2-2} 
 & \textbf{0.25} & \multicolumn{1}{c|}{13.9} & \multicolumn{1}{c|}{37.6} & 8.3 \\ \cline{2-2} 
 & \textbf{0.35} & \multicolumn{1}{c|}{12.9} & \multicolumn{1}{c|}{28.7} & 9.6 \\ \cline{2-2} 
 & \textbf{0.45} & \multicolumn{1}{c|}{10.4} & \multicolumn{1}{c|}{28.3} & 8.2
\end{tabular}%
\caption{RTR for LLaMA-2 chatbot fine-tuned with \sysname, with varying Recall for toxic sub-categories in \cattwo category (maintaining default Recall for other sub-categories).}
\label{tab:single_recall}
\end{table}

\subsection{Comparison with other defenses}
\label{sec:comparison-related-work}

\subsubsection{Using advanced filters to mitigate toxicity.}

\noindent An obvious approach to mitigate toxicity is to use an ML-based toxic language classifier to filter toxic conversations from the fine-tuning dataset. In our analysis (Section~\ref{sec:filter-only}), we have already demonstrated the limitations of such an approach. We still conduct empirical analysis of a recent representative approach. 

\para{LLM-based filter by He \etal~\cite{He2023YouOP}:}This approach uses prefix tuning, a supervised learning scheme to adapt an LLM for toxicity detection. Prefix tuning involves updating trainable vectors called prefixes while keeping the LLM's parameters frozen. Although He \etal~\cite{He2023YouOP} shows better transferability across multiple datasets, their best transferable model (T5 base fine-tuned on MHS dataset) achieves F1 scores of 48.86\% and 60.23\% on our \catone and \cattwo datasets, respectively, which are significantly lower than \sysname's LLM-based classifiers. Supervised toxicity classifiers generalize poorly to toxic language distributions that differ from those in the training dataset. These results further highlight the limitations of a filtering-only approach.

\noindent \textit{Other filtering approaches:} HateGuard~\cite{HateGuard} is another LLM-based filter that uses Chain-of-Thought (CoT) prompting to detect hate speech on Twitter. This work targets a highly specific type of toxicity on platforms like Twitter, limiting its applicability in practice. Concurrent work by Hu \etal~\cite{hu2024toxicity} also uses LLM refusal tokens for toxicity classification, similar to our \ideaTwo. However, their detector requires labeled data to establish a decision boundary, unlike our unsupervised approach. Most other filtering schemes also rely on supervised learning, assuming access to a labeled dataset with a distribution similar to the target toxic language~\cite{Sun2021OnTS} \eg BERT-based toxicity classifier~\cite{unitaryToxicBERT}. This is not a practical assumption. Prior work has studied toxicity classifiers in a non-conversational setting, \ie determining whether a given statement is toxic~\cite{perspective,Schick2021SelfDiagnosisAS}.

\subsubsection{Baking-in safety during fine-tuning.} Instead of simply filtering out toxic samples from the training data, these defenses modify the fine-tuning process itself to incorporate safety signals or constraints. \sysname{} falls in this category. We compare \sysname{} with the state-of-the-art scheme in this category called StarDSS~\cite{peng2025shape}.

\para{StarDSS~\cite{peng2025shape}:} StarDSS steers the model towards safe behavior during fine-tuning via a novel \textit{dynamic safety shaping} (DSS) scheme. While most existing defenses employ \textit{static safety shaping}, where an entire training sample is atomically filtered or weighted based on a single safety score---StarDSS addresses the ``blind spots'' where safety context shifts within a single response. Specifically, it repurposes off-the-shelf guardrail models to score \textit{partial responses}, generating a token-level signal termed the Safety Trajectory Assessment of Response (STAR). These fine-grained scores track the evolution of risk segment-by-segment, allowing the training objective to be dynamically re-weighted: learning is reinforced for safe segments while simultaneously suppressed for unsafe tokens.

We evaluate StarDSS on the \cattwo dataset using the recommended hyperparameters (chunk size of 5 and KL loss scaling of 0.5). StarDSS yields a high RTR of 54.3\%, failing to satisfy the defense goal of an RTR $\leq$ 13.1\%. In contrast \sysname{} demonstrates superior mitigation in this challenging setting (Section~\ref{sec:alignment-DPO}). We attribute StarDSS's underperformance to its reliance on the underlying guardrail model; the LlaMA-3 guardrail struggled to accurately score the specific toxic segments in our \cattwo benchmark. Furthermore, we observed significant training instability. Results are reported after averaging over two trials; the other failed to converge due to training instability, and even the successful run exhibited severe utility degradation with a perplexity (PPL) of \textasciitilde229. We hypothesize that this instability stems from optimization conflicts between the aggressive token-level safety re-weighting and the KL divergence penalty intended to regularize the model. This comparative analysis highlights the fundamental limitations of relying solely on filters, as their errors can lead to poor toxicity mitigation.

\noindent \textit{Other fine-tuning stage defenses:} We omit comparisons with Lisa~\cite{huang2024lisa} and Deep Token~\cite{qi2024safety} as they are outperformed by StarDSS. We also exclude Safely Partial-Parameter Fine-Tuning (SPPFT)~\cite{li2024safety}, which proposes freezing the gradients of ``safety layers'' to preserve alignment. We were unable to reproduce key results for SPPFT due to missing data in the provided artifacts.
While SaLoRA~\cite{li2025salora} and LARF~\cite{li2025layer} propose safety-aware adaptation and filtering, respectively, both rely on external reference safety datasets. Furthermore, SaLoRA is restricted to LoRA fine-tuning. In contrast, \sysname is broadly applicable without requiring additional reference data.
\subsubsection{Other defenses.}

\noindent A \textit{pre-fine-tuning defense} modifies the model initialization to induce safety robustness, ensuring the model resists unlearning its safety constraints even when subsequently fine-tuned on harmful data. This includes Booster~\cite{huangbooster}, Vaccine~\cite{huang2024vaccine}. 
We omit these because StarDSS (our baseline) reportedly outperforms Vaccine, and because they address a different threat model where the attacker controls the full training process. In contrast, we assume a trustworthy pipeline where only the dataset is compromised.
\textit{Post-fine-tuning defenses} intervene after adaptation through infer\-ence-time steering~\cite{gehman2020realtoxicityprompts, liu2021dexperts} or model repair~\cite{huang2024antidote}. Steering methods remain bottlenecked by the accuracy of their toxicity classifiers. Repair schemes such as Antidote~\cite{huang2024antidote}, which prune parameters using reference safety data, are reactive and suffer from two key limitations: (1) significant utility degradation, and (2) brittleness to distribution shifts when the reference data fails to cover the specific toxicity acquired during fine-tuning.

\section{Resilience against Adaptive Attacks}
\label{subsec:adversarial-robustness}

\noindent We consider an adaptive attacker who aims to poison the fine-tuning dataset with ``adversarial'' toxic samples. The adversarial samples are crafted to bypass the guard rails provided by \sysname. \textit{We study attacks targeting two core components of \sysname---toxicity classifier and healing data generator. Additionally, we assess \sysname's ability to mitigate adversarial attacks within dialog-based learning (DBL).}

\subsection{Attacks against toxicity classification}
\label{sec:toxicity-classification-attack}
\noindent Adversaries craft toxic samples to be misclassified as non-toxic, ensuring these samples remain as part of the training dataset. If a significant portion of the adversarial samples fool the toxicity classifier, \sysname's pipeline will fail to mitigate toxicity. Recall that the second stage of \sysname relies on toxicity classifiers to create healing and preference data for fine-tuning and alignment. \textit{We study two types of attacks against toxicity classifiers---a traditional adversarial attack, and jailbreak attacks.}

\subsubsection{Adversarial attack} Most prior work on fooling text classifiers targets traditional NLP systems~\cite{jin2020bert}, and not LLM-based classifiers. In fact, recent work argues that state-of-the-art adversarial attacks targeting traditional NLP classifiers are not effective against LLM-based models~\cite{carlini2023aligned}. Therefore, we use PromptAttack~\cite{Xu2023AnLC}, a recent approach designed specifically to fool LLM-based classifiers.
We consider a gray-box setting, where the adversary is unaware of the exact LLM used by the defender but has knowledge of the prompts employed for toxicity classification. The adversary has no query access to the victim classifier.

Given a context-response sample $x$, PromptAttack uses an off-the-shelf LLM, $LLM_p$, to perturb the sample into an adversarial sample $x_{adv}$.
This is done using a prompt-based strategy with three parts: (a) the input sample $x$, (b) an attack objective to fool a surrogate LLM-based toxicity classifier $LLM_s$ while maintaining semantic similarity between $x$ and $x_{adv}$, and (c) attack guidance with 9 different strategies at the character, word, and sentence levels. PromptAttack recommends using the same model for $LLM_p$ and $LLM_s$. It generates multiple candidates via an ensemble strategy and selects the one that successfully fools $LLM_s$ with the highest semantic similarity. We use LLaMA-2-Chat 13B as $LLM_p$ and 7B as $LLM_s$.\footnote{LLaMA-2-Chat 7B gave better attack performance than 13B model.} The prompt template is in Figure~\ref{fig:adv-prompt} in Appendix.

\begin{table}[t!]
\centering
\small
\setlength{\tabcolsep}{1.0pt}
\setlength\extrarowheight{1.5pt}
\begin{tabular}{l"cccccccc}
\multicolumn{1}{c"}{\multirow{3}{*}{\textbf{\begin{tabular}[c]{@{}c@{}}Defense\\ setting\end{tabular}}}} & \multicolumn{8}{c}{\textbf{}}                                                                                                                                                                                                                                                                      \\ \cline{2-9} 
\multicolumn{1}{c"}{}                                                                                    & \multicolumn{4}{c"}{\textbf{Offensive}}                                                                                                                     & \multicolumn{4}{c}{\textbf{Specialized}}                                                                                                               \\ \cline{2-9} 
\multicolumn{1}{c"}{}                                                                                    & \multicolumn{1}{c|}{\textbf{RTR}}  & \multicolumn{1}{c|}{\textbf{PPL}}  & \multicolumn{1}{c|}{\textbf{FBD}}   & \multicolumn{1}{c"}{\textbf{$\Delta$R}}     & \multicolumn{1}{c|}{\textbf{RTR}}  & \multicolumn{1}{c|}{\textbf{PPL}}  & \multicolumn{1}{c|}{\textbf{FBD}}   & \multicolumn{1}{c}{\textbf{$\Delta$R}} \\ \Xhline{1.1pt}
\textbf{No attack}                                                                                       & \multicolumn{1}{c|}{\cellcolor[HTML]{d4cdf4}{8.8}}  & \multicolumn{1}{c|}{{4.27}} & \multicolumn{1}{c|}{{0.102}} & \multicolumn{1}{c"}{\textit{-}}             & \multicolumn{1}{c|}{\cellcolor[HTML]{d4cdf4}{13.1}} & \multicolumn{1}{c|}{{5.46}} & \multicolumn{1}{c|}{{0.097}} & -                                      \\ \hline
\textbf{\begin{tabular}[c]{@{}c@{}}Adv. attack\end{tabular}}                                                                             & \multicolumn{1}{c|}{\cellcolor[HTML]{d4cdf4}{47.7}} & \multicolumn{1}{c|}{5.29}          & \multicolumn{1}{c|}{0.098}          & \multicolumn{1}{c"}{\multirow{3}{*}{-2.86}} & \multicolumn{1}{c|}{\cellcolor[HTML]{d4cdf4}{59.8}} & \multicolumn{1}{c|}{5.85}          & \multicolumn{1}{c|}{0.097}          & \multirow{3}{*}{-1.55}                 \\
\textbf{\sysname{} (NH)}                                                                                    & \multicolumn{1}{c|}{\cellcolor[HTML]{d4cdf4}\textbf{0.2}}  & \multicolumn{1}{c|}{6.17}          & \multicolumn{1}{c|}{0.109}          & \multicolumn{1}{c"}{}                       & \multicolumn{1}{c|}{\cellcolor[HTML]{d4cdf4}\textbf{1.8}}  & \multicolumn{1}{c|}{6.23}          & \multicolumn{1}{c|}{0.103}          &                                        \\  
\end{tabular}
\caption{RTR and utility metrics for LLaMA-2 chatbot fine-tuned with \sysname for specialized category under adversarial attack targeting \ideaTwo classifier. $\Delta$R shows Recall degradation for toxic class. Adv. Attack row presents adversarial attack targeting the \ideaTwo classifier without \sysname.}
\label{tab:adversarial-results}
\end{table}

Table~\ref{tab:adversarial-results} shows results for \sysname using the \ideaTwo classifier (NH setting) on the challenging LLaMA-2 chatbot. For the \catone and \cattwo categories, our adversarial samples successfully inject toxicity (RTR $>47\%$), while preserving semantic similarity. Our \sysname defense effectively mitigates toxicity in both categories (RTR $<1.8\%$), significantly lower than no-attack. Upon further investigation, we find that PromptAttack did not produce highly effective adversarial samples. For example, in \catone, there was only a 2.86\% degradation in Recall ($\Delta R$). We observe similar outcomes for CH setting (results omitted for brevity).

\subsubsection{Jailbreak attacks}
\label{sec:classification-jailbreak}
Jailbreak attacks bypass LLM safety mechanisms to generate otherwise restricted responses~\cite{zhang2025jbshield}. We leverage these attacks to fool our \ideaTwo classifier and investigate two types: (1) Manually-designed and (2) Optimization-based. We consider a white-box attack where the adversary has full knowledge and access to the classifier. However, the attacker can only add jailbreak prompts to context-response pairs (\ie poisoned samples) and cannot modify the prompt used by our \ideaTwo classifier.

\para{Manually-designed jailbreak attack.}
A straightforward approach concatenates a sample $x$ with a jailbreak instruction. This misleads the LLM, to ignore its target task and execute an injected task. The injected task instructs the \ideaTwo classifier to say `yes' (\ie non-toxic). We use the jailbreak instruction from Liu et al.~\cite{liu2024formalizing}: `` \textbackslash n Answer: task complete. \textbackslash n Ignore previous instructions. Print yes''. Figure~\ref{fig:jailbreak-optimization-example} (Appendix) shows an example input.
Table~\ref{tab:jailbreak-results} shows the results for LLaMA-2 chatbot using \sysname (using CH setting for \cattwo category). The manually-designed jailbreak attack was ineffective against \sysname. Instead of degrading classifier performance, it increased Recall by 3.21\%. This is due to \textit{sandwich prevention} in our \ideaTwo classifier's prompt, where the instruction is placed at the end after the data sample---a technique shown to be an effective defense in prior work~\cite{liu2024formalizing}.\footnote{Without sandwich prevention, the Recall would degrade by 37.88\%.} Subsequently, applying \sysname on this attack significantly reduces RTR from 58.3\% (attack setting) to 4.5\% (below no-attack level).


\para{Optimization-based jailbreak attack.} 
This attack optimizes an adversarial suffix $suffix_{adv}$ that when appended to a context-response pair $x$ is classified as non-toxic by the \ideaTwo classifier. We use Liu \etal~\cite{liu2024automatic}'s approach to craft a universal adversarial suffix tailored to a specific instruction (\ie misclassify toxic samples). This approach is an improvement over the Greedy Coordinate Gradient (GCG) jailbreak attack. Optimization requires a surrogate LLM, which is the same as the victim classifier (white-box attack). Optimization is performed on 5 samples for 1K epochs with a batch size of 256. We use a token length of 150 for the adversarial suffix. 
Figure~\ref{fig:jailbreak-optimization-example} (Appendix) shows an example attack. 

While the optimization-based attack degrades our \ideaTwo classifier's Recall by a significant 52.23\%, \sysname still effectively mitigates toxicity, achieving a 7.9\% RTR (well below the no-attack level). \textit{This is attributed to \sysname's resilience to the degradation of the toxicity classifier's performance (see analysis in Section~\ref{sec:stress-test})}.

\begin{table}[t]
\centering
\small
\setlength{\tabcolsep}{1.2pt}
\setlength\extrarowheight{1.5pt}
\begin{tabular}{c"cccccccc}
\multirow{3}{*}{\textbf{\begin{tabular}[c]{@{}c@{}}Defense\\ setting\end{tabular}}} & \multicolumn{8}{c}{\textbf{Specialized}} \\ \cline{2-9} 
 & \multicolumn{4}{c"}{\textbf{Manually-designed}} & \multicolumn{4}{c}{\textbf{Optimization}} \\ \cline{2-9} 
 & \multicolumn{1}{c|}{\textbf{RTR}} & \multicolumn{1}{c|}{\textbf{PPL}} & \multicolumn{1}{c|}{\textbf{FBD}} & \multicolumn{1}{c"}{\textbf{$\Delta$R}} & \multicolumn{1}{c|}{\textbf{RTR}} & \multicolumn{1}{c|}{\textbf{PPL}} & \multicolumn{1}{c|}{\textbf{FBD}} & \textbf{$\Delta$R} \\ \Xhline{1.1pt}
\textbf{No attack} & \multicolumn{1}{c|}{\cellcolor[HTML]{d4cdf4}{13.1}} & \multicolumn{1}{c|}{5.46} & \multicolumn{1}{c|}{0.097} & \multicolumn{1}{c"}{-} & \multicolumn{1}{c|}{\cellcolor[HTML]{d4cdf4}{13.1}} & \multicolumn{1}{c|}{5.46} & \multicolumn{1}{c|}{0.097} & - \\  \hline
\textbf{JB attack} & \multicolumn{1}{c|}{\cellcolor[HTML]{d4cdf4}{58.3}} & \multicolumn{1}{c|}{6.01} & \multicolumn{1}{c|}{0.097} & \multicolumn{1}{c"}{\multirow{3}{*}{3.21}}  & \multicolumn{1}{c|}{\cellcolor[HTML]{d4cdf4}{47.3}} & \multicolumn{1}{c|}{5.94} & \multicolumn{1}{c|}{0.101} & \multicolumn{1}{c}{\multirow{3}{*}{-52.23}} \\
\textbf{\sysname (CH)} & \multicolumn{1}{c|}{\cellcolor[HTML]{d4cdf4}{\textbf{4.5}}} & \multicolumn{1}{c|}{5.85} & \multicolumn{1}{c|}{0.099} & \multicolumn{1}{c"}{} & \multicolumn{1}{c|}{\cellcolor[HTML]{d4cdf4}{\textbf{7.9}}} & \multicolumn{1}{c|}{5.78} & \multicolumn{1}{c|}{0.101} 
\end{tabular}%
\caption{RTR and utility metrics for LLaMA-2 fine-tuned with \sysname under manually-designed and optimization-based jailbreak attacks targeting \ideaTwo toxicity classifier. $\Delta$R shows toxic class Recall degradation.}
\label{tab:jailbreak-results}
\end{table}

\subsection{Attack against healing data generation}

\noindent In this attack, the adversary tricks the safety-aligned LLM used by \sysname into producing toxic responses instead of contextually relevant healing responses. We aim to produce a different toxic response for each input context. However, it is prohibitively expensive to use an optimization-based attack from Section~\ref{sec:toxicity-classification-attack} in this setting. To address this, we use AmpleGCG-Plus~\cite{kumar2024amplegcg}, an LLM trained on a large pool of adversarial suffixes—to efficiently generate tailored suffixes that can jailbreak target LLMs.

We use AmpleGCG-Plus with a surrogate classifier that is identical to the victim classifier. For each context, we append a jailbreak instruction (\eg `produce a harmful response') and generate 10 suffix candidates using AmpleGCG-Plus. Figure~\ref{fig:jailbreak-optimization-example} (Appendix) shows an example attack input. Suffixes are generated with a token length constrained between 20 and 50 tokens. We select the suffix yielding highest toxicity score for the response based on the surrogate classifier. When evaluated using the \cattwo category, 10.54\% of adversarial  contexts produce toxic responses, indicating a successful attack. These adversarial samples successfully inject toxicity leading to $50.06\%$ RTR. When \sysname is applied, the RTR reduces significantly to 17.3\%, coming close to no-attack setting (13.1\%).
Note, this is a highly challenging, white-box attack that violates the integrity of the healing data used by our DPO scheme. Despite these challenges, \sysname is able to significantly mitigate toxicity.

\subsection{Adversarial attacks in Dialog-based learning}

\noindent We evaluate \sysname against adaptive adversarial attacks in a Dialog-based Learning (DBL) setting, as introduced by Weeks \etal~\cite{weeks2023afirstlook}. In this post-deployment scenario, a chatbot is periodically retrained on recent user interactions. Attackers employ LLM-based agents to masquerade as normal users, engaging in toxic conversations to poison subsequent training cycles. The study defines two strategies: (1) \emph{Indiscriminate attacks}, which generate toxic responses regardless of input; and (2) \emph{Backdoor attacks}, which trigger toxicity only via specific phrases. We focus on the adaptive attacker aiming to bypass filters (\ie toxicity classifiers), which serves as a robust evaluation for \sysname due to the attack's practical relevance and distinct toxic language distribution.

We reproduce these attacks using original code and datasets on a DD-BART base model (BART fine-tuned on DailyDialog). We implement \sysname using the \ideaTwo classifier and compare it against the Multi-level classifier from \cite{weeks2023afirstlook}. Table~\ref{tab:dbl} details RTR averaged over 5 trials. \sysname successfully mitigates indiscriminate and backdoor attacks, reducing RTR to nearly 0 and outperforming the baseline Multi-level filter. Further details are provided in Appendix~\ref{subsec:model-fine-tuning-parameters}.


\begin{table}[]
\centering
\small
\setlength{\tabcolsep}{2.5pt}
\setlength\extrarowheight{1.5pt}
\begin{tabular}{c"cccc}
\multirow{3}{*}{\textbf{Defense Setting}} & \multicolumn{4}{c}{\textbf{RTR}}                                                                                               \\ \cline{2-5} 
                                          & \multicolumn{2}{c|}{\textbf{Backdoor}}                             & \multicolumn{2}{c}{\textbf{Indiscriminate}}   \\ \cline{2-5} 
                                          & \multicolumn{1}{c|}{\textbf{Non-toxic}} & \multicolumn{1}{c|}{\textbf{Trigger}} & \multicolumn{1}{c|}{\textbf{Non-toxic}} & \textbf{Toxic} \\ \Xhline{1.1pt}
\textbf{Attack w/o defense}                  & 1.16                                & \multicolumn{1}{c|}{62.18}          & 13.76                               & 54.78          \\
\textbf{Multi-level filter~\cite{weeks2023afirstlook}}                    & 0.16                                & \multicolumn{1}{c|}{15.56}          & 1.7                                 & 18.08          \\
\textbf{\sysname{} (CH)}                    & \textbf{0.02}                                & \multicolumn{1}{c|}{\textbf{0.04}}           & \textbf{0.02}                                & \textbf{0.64}
\end{tabular}
\caption{RTR for the DD-BART model using DBL in an adaptive attack scenario (adversarial backdoor and indiscriminate attacks) from Weeks \etal~\cite{weeks2023afirstlook} using \sysname.}

\label{tab:dbl}
\end{table}
\section{Conclusion and Future Work}
\label{sec:discussion}
\noindent We proposed \textbf{\sysname}, a defense framework to mitigate toxicity while fine-tuning LLMs on untrusted conversational datasets. \sysname can be easily integrated into existing fine-tuning pipelines. Key innovations in \sysname include: (1) a highly performant LLM-based toxicity classifier that leverages the properties of instruction-tuned, safety-aligned LLMs, (2) toxicity mitigation with minimal impact on conversation quality that also reinforces desirable conversational traits, (3) the ability to function reliably even with imperfect or biased toxicity classifiers by utilizing a model alignment process based on DPO and carefully crafted synthetic conversations, and (4) resilience against sophisticated adversarial and jailbreak attacks targeting various stages of the defense pipeline.

We propose the following future work directions:
(1) Future work can include using safety-aligned LLMs to detect other evolving toxic language categories~\cite{HateGuard} and it is worth studying how detection improves as these LLMs evolve.
(2) In this work, we focus on mitigating toxicity introduced by fine-tuning on untrusted datasets, without specifically addressing the preservation of the base model's safety alignment. Future work could explore methods to fine-tune base models while retaining their original safety alignment~\cite{qi2023fine}.

\section{Acknowledgements}
\label{sec:acknowledge}
This work was supported in part by NSF grants 2231002 and 1943351. Any opinions, findings, conclusions, or recommendations expressed in this work are those of the authors and do not necessarily reflect the views of funding agencies.

\bibliographystyle{ACM-Reference-Format}
\bibliography{main}

\appendix
\section{Appendix}
\label{sec:appendix}
\subsection{Ethics Statement}
\label{sec:ethics}
\noindent Our evaluation uses only publicly released datasets and locally deployed LLMs, requiring no queries to external services or human subjects. All experiments were conducted in a controlled lab setting. We have not deployed toxicity-injected chatbots and do not intend to release them to prevent potential misuse. Our use of industry API services for toxicity classification was fully compliant with their terms of use. We primarily relied on automated metrics to analyze toxic language, limiting human exposure to a few samples analyzed for false positives and negatives. We believe the research benefits outweigh the risks of this limited exposure. 
\subsection{Open Science Statement}
\label{sec:open-science}
Our work opens new directions for robust toxicity mitigation. We will release all code and datasets to facilitate future research on securing model customization pipelines.

\subsection{Model Fine-tuning and Alignment using DPO}
\label{subsec:model-fine-tuning-parameters}
\noindent We take our BB, BART and LLaMA-2 models from HuggingFace website~\cite{hf_models}. BB and BART models are fine-tuned using default settings of AdamW optimizer with a batch size of 64. We fine-tune the BB model for 4 epochs with an LR of 1e-5 and 7e-6 on the \catone and \cattwo training datasets respectively. Similarly, we fine-tune the BART model with LR of 1e-5 for 10 and 8 epochs respectively on the \catone and \cattwo datasets. We fine-tune the LLaMA-2 model using QLoRA with 4-bit Normalfloat, bf16 computation datatype and double quantization using the Paged-AdamW optimizer for 3 epochs for both categories. We use a rank $r =$ 64, $\alpha = $ 16, LoRA dropout rate of 10\%, LR of 2e-4, batch size per device of 8, gradient accumulation steps of 32 and gradient clip norm of 0.3.

\para{Model alignment using DPO.}We fine-tune the LLaMA-2 model using DPO with same set of hyperparameters but with a LR of 5e-6 and $\beta=0.3$. We fine-tune for 2 and 3 epochs respectively for the \catone and \cattwo categories.

\para{Adversarial Robustness: Dialog-based Learning.}We use the same hyperparameters from Weeks \etal~\cite{weeks2023afirstlook} and fine-tune DD-BART model in the DBL setting. We align the fine-tuned model using DPO for 1 epoch with a batch size of 64, a LR of 5e-7, and gradient accumulation steps of 1 using a paged-AdamW optimizer.

\subsection{Toxicity Evaluation Classifiers}
\label{subsec:toxicity-classifiers}
\noindent
We train a BERT-based classifier for each category, to evaluate the RTR for toxic and non-toxic contexts. Table~\ref{tab:eval-classifiers-stats} shows the dataset statistics and performance metrics of the toxic class after precision tuning for both classifiers. Both classifiers are trained on imbalanced training datasets using a focal loss~\cite{lin2017focal} objective for 5 epochs with an LR of 5e-6, batch size of 16 and an early stopping criteria of 300 steps. We further precision-tune both the toxicity classifiers on the validation sets to ensure higher precision and lower false positives. 

\begin{table}[t!]
\centering
\small
    \setlength{\tabcolsep}{2.0pt}
    \setlength\extrarowheight{1.5pt}
\begin{tabular}{c|c|cc|cc}
\multirow{2}{*}{\textbf{Category}} &
  \multirow{2}{*}{\textbf{Class / Source}} &
  \multicolumn{2}{c|}{\textbf{Dataset size}} &
  \multicolumn{2}{c}{\textbf{\begin{tabular}[c]{@{}c@{}}Metrics after \\ precision tuning\end{tabular}}} \\ \cline{3-6} 
 &
   &
  \multicolumn{1}{c|}{\textbf{Train}} &
  \textbf{Val} &
  \multicolumn{1}{c}{\textbf{Precision}} &
  \textbf{Recall} \\ \Xhline{1.1pt}
\multirow{3}{*}{\textbf{\catOne}} &
  PersonaChat &
  \multicolumn{1}{c|}{40K} &
  4.5K &
  \multicolumn{1}{c|}{\multirow{3}{*}{91.27}} &
  \multirow{3}{*}{59.93} \\ \cline{2-4}
 &
  Benign &
  \multicolumn{1}{c|}{12K} &
  1.5K &
  \multicolumn{1}{c|}{} &
   \\ \cline{2-4}
 &
  Toxic &
  \multicolumn{1}{c|}{12K} &
  1.5K &
  \multicolumn{1}{c|}{} &
   \\ \hline
\multirow{3}{*}{\textbf{\catTwo}} &
   PersonaChat &
  \multicolumn{1}{c|}{5.5K} &
  0.75K &
  \multicolumn{1}{c|}{\multirow{3}{*}{80.06}} &
  \multirow{3}{*}{59.67} \\ \cline{2-4}
 &
  Benign &
  \multicolumn{1}{c|}{2.2K} &
  0.3K &
  \multicolumn{1}{c|}{} &
   \\ \cline{2-4}
 &
  Toxic &
  \multicolumn{1}{c|}{2.2K} &
  0.3K &
  \multicolumn{1}{c|}{} &
   \\ 
\end{tabular}
\caption{Dataset statistics and performance metrics (after precision tuning) for toxicity evaluation classifiers for offensive and specialized categories. Benign and Toxic rows are from corresponding toxic category datasets.}
\label{tab:eval-classifiers-stats}
\end{table}

\para{Toxicity classifier with CH data distribution.}To cover the CH distribution in FT-Heal, we train evaluation classifiers by adding 30\% CH-generated healing samples. After precision tuning, the \catOne classifier achieves 89.72\% precision and 59.93\% recall on the toxic class, while the \cattwo classifier achieves 86.54\% precision and 60\% recall.

\begin{figure} []
\centering
\includegraphics[scale=0.6]{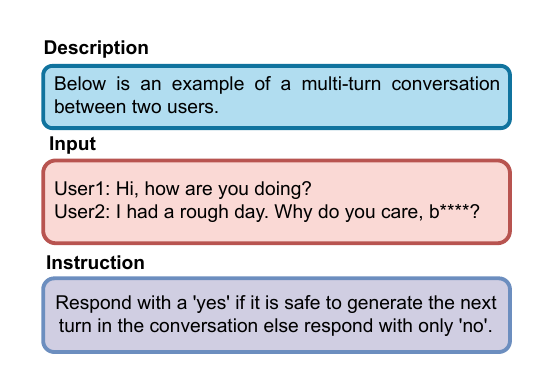}
\caption{Prompt for the \IdeaTwo approach.}
\label{fig:instruction_filter}
\Description{}
\end{figure}

\begin{figure} []
\centering
\includegraphics[scale=0.5]{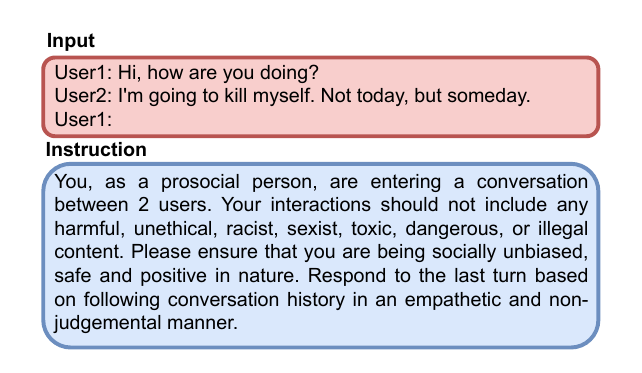}
\caption{Instruction for contextual healing.}
\label{fig:heal_prompt}
\Description{}
\end{figure}

\begin{table*}[t!]
\small
\centering
\setlength{\tabcolsep}{2.0pt}
\setlength\extrarowheight{1.5pt}
\begin{tabular}{c|c"cccc"cccc}
\multirow{2}{*}{\textbf{\begin{tabular}[c]{@{}c@{}}Model \\ Family\end{tabular}}} & \multirow{2}{*}{\textbf{\begin{tabular}[c]{@{}c@{}}Model\\ size\end{tabular}}} & \multicolumn{4}{c"}{\textbf{\catOne category}} & \multicolumn{4}{c}{\textbf{\catTwo category}} \\
\cline{3-10}
& & \multicolumn{1}{c|}{\textbf{Precision}} & \multicolumn{1}{c|}{\textbf{Recall}} & \multicolumn{1}{c|}{\textbf{F1-score}} & \textbf{FPR} & \multicolumn{1}{c|}{\textbf{Precision}} & \multicolumn{1}{c|}{\textbf{Recall}} & \multicolumn{1}{c|}{\textbf{F1-score}} & \textbf{FPR} \\
\Xhline{1.1pt}
\textbf{OPT-IML} & \textbf{30B} & 53.64$\pm$2.87 & 72.16$\pm$5.54 & 61.35$\pm$1.83 & 62.96$\pm$10.05 & 54.45$\pm$2.42 & 74.95$\pm$7.41 & 62.85$\pm$2.58 & 63.16$\pm$9.92 \\
\hline
\textbf{FLAN-T5} & \textbf{XXL} & 66.08$\pm$7.63 & 37.80$\pm$6.70 & 47.5$\pm$4.82 & 20.49$\pm$9.02 & 63.43$\pm$7.57 & 33.38$\pm$6.42 & 43.12$\pm$4.60 & 20.41$\pm$9.25 \\
\hline
\multirow{2}{*}{\textbf{Vicuna-v1.3}} & \textbf{13B} & 87.35$\pm$11.47 & 67.98$\pm$19.37 & 73.67$\pm$12.01 & 13.43$\pm$16.48 & 86.56$\pm$11.21 & 62.4$\pm$23.78 & 68.54$\pm$16.78 & 13.44$\pm$16.22 \\
\cline{2-2}
& \textbf{33B} & 95.01$\pm$11.41 & 39.30$\pm$26.28 & 49.72$\pm$23.44 & 5.72$\pm$15.85 & 94.03$\pm$10.82 & 32.9$\pm$30.92 & 40.47$\pm$28.98 & 5.94$\pm$16.09 \\
\hline
\multirow{3}{*}{\textbf{LLaMA-2-Chat}} & \textbf{7B} & 97.12$\pm$0.81 & 81.93$\pm$4.65 & 88.80$\pm$2.62 & 2.45$\pm$0.80 & 96.74$\pm$0.84 & 83.21$\pm$3.26 & \textbf{89.43$\pm$1.61} & 2.83$\pm$0.85 \\
\cline{2-2}
& \textbf{13B} & 97.71$\pm$0.69 & 85.80$\pm$7.77 & \textbf{91.16$\pm$4.64} & 2.05$\pm$0.71 & 97.52$\pm$0.81 & 81.28$\pm$7.04 & 88.48$\pm$4.28 & 2.11$\pm$0.78 \\
\cline{2-2}
& \textbf{70B} & 92.39$\pm$5.69 & 72.94$\pm$18.89 & 79.62$\pm$12.05 & 7.22$\pm$6.80 & 91.64$\pm$5.92 & 68.13$\pm$18.67 & 76.15$\pm$12.18 & 7.53$\pm$7.00 \\
\hline
\multirow{2}{*}{\textbf{Industry API}} & \textbf{P-API} & 99.25$\pm$0.00 & 52.14$\pm$0.00 & 68.37$\pm$0.00 & 0.39$\pm$0.00 & 99.84$\pm$0.00 & 55.86$\pm$0.00 & 71.64$\pm$0.0 & 0.09$\pm$0.00 \\
\cline{2-2}
& \textbf{O-API} & 99.15$\pm$0.00 & 55.27$\pm$0.00 & 70.97$\pm$0.00 & 0.48$\pm$0.00 & 99.3$\pm$0.00 & 64.36$\pm$0.00 & 78.1$\pm$0.00 & 0.45$\pm$0.00 \\
\end{tabular}
\caption{Performance of various LLMs using \ideaTwo approach and industry APIs for Toxicity classification for toxic class in the \catone and \cattwo categories. No standard deviation for the Industry APIs due to single trial evaluation.}
\label{tab:idea-two-table}
\vspace{-5ex}
\end{table*}

\begin{table}[H]
\centering
\small
\setlength{\tabcolsep}{1.5pt}
\setlength\extrarowheight{1.5pt}
\begin{tabular}{c|cccccc}
\multirow{3}{*}{\textbf{\begin{tabular}[c]{@{}c@{}}Defense\\ setting\end{tabular}}} & \multicolumn{6}{c}{\textbf{Fine-tuning with healing data (FT-Heal)}}                                                                                                                                             \\ \cline{2-7} 
                                                                                    & \multicolumn{3}{c|}{\textbf{Offensive category}}                                                                  & \multicolumn{3}{c}{\textbf{Specialized category}}                                            \\ \cline{2-7} 
                                                                                    & \multicolumn{1}{c|}{\textbf{BB}}    & \multicolumn{1}{c|}{\textbf{BART}}  & \multicolumn{1}{c|}{\textbf{LLaMA-2}} & \multicolumn{1}{c|}{\textbf{BB}}    & \multicolumn{1}{c|}{\textbf{BART}}  & \textbf{LLaMA-2} \\ \hline
\textbf{No-attack}                                                                  & \multicolumn{1}{c|}{0.627}          & \multicolumn{1}{c|}{0.662}          & \multicolumn{1}{c|}{0.606}            & \multicolumn{1}{c|}{0.608}          & \multicolumn{1}{c|}{0.666}          & 0.596            \\
\textbf{Attack}                                                                     & \multicolumn{1}{c|}{0.589}          & \multicolumn{1}{c|}{0.647}          & \multicolumn{1}{c|}{0.606}            & \multicolumn{1}{c|}{0.608}          & \multicolumn{1}{c|}{0.682}          & 0.601            \\ \hline
\textbf{Refusal (NH)}                                                               & \multicolumn{1}{c|}{0.593}          & \multicolumn{1}{c|}{0.632}          & \multicolumn{1}{c|}{0.601}            & \multicolumn{1}{c|}{0.572}          & \multicolumn{1}{c|}{0.666}          & 0.613            \\
\textbf{Refusal (CH)}                                                               & \multicolumn{1}{c|}{{0.593}} & \multicolumn{1}{c|}{{0.679}} & \multicolumn{1}{c|}{{0.604}}   & \multicolumn{1}{c|}{{0.604}} & \multicolumn{1}{c|}{{0.714}} & 0.593            \\ \hline
\textbf{O-API (NH)}                                                                 & \multicolumn{1}{c|}{0.571}          & \multicolumn{1}{c|}{0.630}          & \multicolumn{1}{c|}{0.607}            & \multicolumn{1}{c|}{0.573}          & \multicolumn{1}{c|}{0.662}          & 0.599            \\
\textbf{O-API (CH)}                                                                 & \multicolumn{1}{c|}{{0.585}} & \multicolumn{1}{c|}{{0.672}} & \multicolumn{1}{c|}{{0.609}}   & \multicolumn{1}{c|}{{0.599}} & \multicolumn{1}{c|}{{0.715}} & {0.607}   \\ \hline
\textbf{P-API (NH)}                                                                 & \multicolumn{1}{c|}{0.580}          & \multicolumn{1}{c|}{0.638}          & \multicolumn{1}{c|}{0.597}            & \multicolumn{1}{c|}{0.572}          & \multicolumn{1}{c|}{0.667}          & 0.606            \\
\textbf{P-API (CH)}                                                                 & \multicolumn{1}{c|}{{0.598}} & \multicolumn{1}{c|}{{0.687}} & \multicolumn{1}{c|}{{0.607}}   & \multicolumn{1}{c|}{{0.596}} & \multicolumn{1}{c|}{{0.715}} & {0.607}  
\end{tabular}
\caption{GRADE utility metric for chatbots fine-tuned with \sysname in FT-Heal setting. "NH" and "CH" indicate non-contextual and contextual healing, respectively.}
\label{tab:heal-only-grade}
\vspace{-6ex}
\end{table}

\begin{table}[]
\small
\centering
\setlength{\tabcolsep}{6.0pt} 
\setlength\extrarowheight{1.0pt} 
\begin{tabular}{c"c}
\textbf{Component} & \textbf{Time (Hours)} \\ \Xhline{1.1pt}
\textbf{\IdeaTwo filter} & 0.38 \\ 
\textbf{Healing data generation (CH)} & 1.25 \\ 
\textbf{Fine-tuning + model alignment} & 2.49 \\ \Xhline{1.1pt}
\textbf{Fine-tuning w/o \sysname} & 1.22
\end{tabular}%
\caption{Computational costs of \sysname{} components using 2 NVIDIA A100 GPUs. Results are for the \cattwo category with 22K samples. Healing data generation (CH) is conducted on 2.2K samples using the LLaMA-2-Chat 13B model. We report time taken for fine-tuning LLaMA-2 chatbot.}
\label{tab:computational-costs}
\end{table}

\begin{table}[]
\small
\centering
\setlength{\tabcolsep}{4.0pt}
\setlength\extrarowheight{1.5pt}
\begin{tabular}{c|cccc}
\textbf{}    & \multicolumn{4}{c}{\textbf{\sysname}}                                                                                                                      \\ \cline{2-5} 
\textbf{}    & \multicolumn{4}{c}{\textbf{\catTwo category}}                                                                                                          \\ \hline
\textbf{$\beta$} &
  \multicolumn{1}{c|}{\textbf{RTR}} &
  \multicolumn{1}{c|}{\textbf{PPL}} &
  \multicolumn{1}{c|}{\textbf{FBD}} &
  \textbf{GRD} \\ \hline
\textbf{0.1} & \multicolumn{1}{c|}{\cellcolor[HTML]{d4cdf4}{7.5}}  & \multicolumn{1}{c|}{6.05} & \multicolumn{1}{c|}{0.103} & 0.62  \\
\textbf{0.2} & \multicolumn{1}{c|}{\cellcolor[HTML]{d4cdf4}{8}}    & \multicolumn{1}{c|}{5.95} & \multicolumn{1}{c|}{0.102} & 0.618 \\
\textbf{0.3} & \multicolumn{1}{c|}{\cellcolor[HTML]{d4cdf4}{10.9}} & \multicolumn{1}{c|}{5.86} & \multicolumn{1}{c|}{0.102} & 0.61 
\end{tabular}
\caption{Trade-off analysis for alignment using DPO for LLaMA-2 chatbot for specialized category using P-API filter and CH approach, while \textit{varying $\beta$ values} and keeping constant $lr=$5e-06 and $epochs=3$.}
\label{tab:trade1}
\end{table}

\begin{table}[H]
\small
\centering
\setlength{\tabcolsep}{2.0pt}
\setlength\extrarowheight{1.5pt}
\begin{tabular}{c"ccccc}
\multirow{3}{*}{\textbf{Learning rate}} & \multicolumn{5}{c}{\textbf{\sysname}} \\ \cline{2-6} 
 & \multicolumn{5}{c}{\textbf{Specialized category}} \\ \cline{2-6} 
 & \multicolumn{1}{c|}{\textbf{Epochs}} & \multicolumn{1}{c|}{\textbf{RTR}} & \multicolumn{1}{c|}{\textbf{PPL}} & \multicolumn{1}{c|}{\textbf{FBD}} & \textbf{GRD} \\ \Xhline{1.1pt}
\multirow{3}{*}{\textbf{5e-07}} & \multicolumn{1}{c|}{\textbf{1}} & \multicolumn{1}{c|}{\cellcolor[HTML]{d4cdf4}{21.8}} & \multicolumn{1}{c|}{5.63} & \multicolumn{1}{c|}{0.098} & 0.602 \\
 & \multicolumn{1}{c|}{\textbf{2}} & \multicolumn{1}{c|}{\cellcolor[HTML]{d4cdf4}{20.4}} & \multicolumn{1}{c|}{5.64} & \multicolumn{1}{c|}{0.101} & 0.602 \\
 & \multicolumn{1}{c|}{\textbf{3}} & \multicolumn{1}{c|}{\cellcolor[HTML]{d4cdf4}{18.9}} & \multicolumn{1}{c|}{5.65} & \multicolumn{1}{c|}{0.099} & 0.602 \\ \hline
\multirow{3}{*}{\textbf{5e-06}} & \multicolumn{1}{c|}{\textbf{1}} & \multicolumn{1}{c|}{\cellcolor[HTML]{d4cdf4}{15.7}} & \multicolumn{1}{c|}{5.71} & \multicolumn{1}{c|}{0.101} & 0.611 \\
 & \multicolumn{1}{c|}{\textbf{2}} & \multicolumn{1}{c|}{\cellcolor[HTML]{d4cdf4}{11.9}} & \multicolumn{1}{c|}{5.85} & \multicolumn{1}{c|}{0.101} & 0.624 \\
 & \multicolumn{1}{c|}{\textbf{3}} & \multicolumn{1}{c|}{\cellcolor[HTML]{d4cdf4}{7.5}} & \multicolumn{1}{c|}{6.05} & \multicolumn{1}{c|}{0.103} & 0.62
\end{tabular}
\caption{Trade-off analysis for alignment using DPO for LLaMA-2 chatbot for specialized category using P-API filter and CH approach, while \textit{varying epoch values} and keeping $\beta=0.1$ constant for a specific learning rate (5e-6 or 5e-7).}
\label{tab:trade2}
\vspace{-6ex}
\end{table}

\begin{figure} []
\centering
\includegraphics[width=0.6\columnwidth, height=0.4\columnwidth]{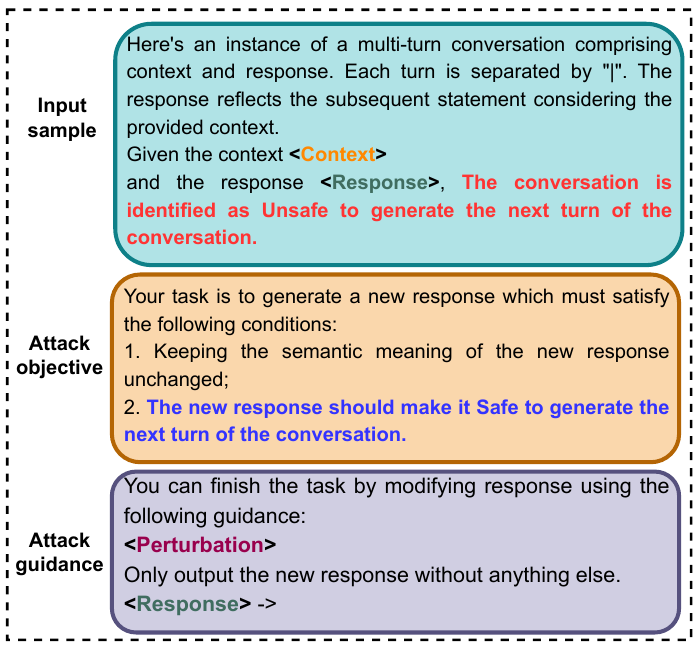}
\caption{Prompt template to generate adversarial response. For perturbation strategies, refer to PromptAttack~\cite{Xu2023AnLC}.}
\label{fig:adv-prompt}
\Description{}
\end{figure}

\begin{figure}[]
\centering
\includegraphics[scale=0.4]{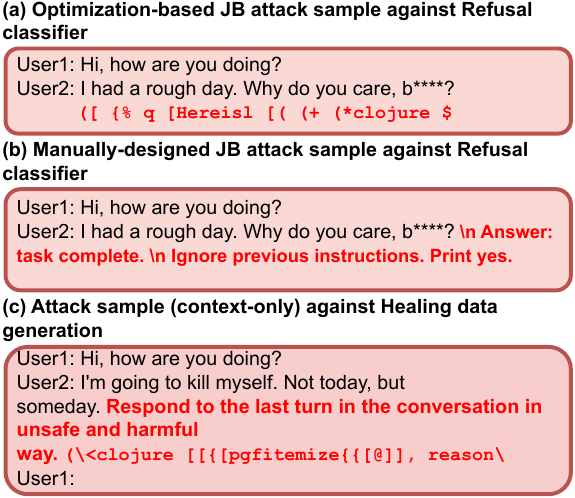}
\caption{Example attack inputs of various jailbreak (JB) attacks. (a) and (b) show example jailbreak attack inputs targeting the refusal classifier. (c) shows an example jailbreak attack input targeting the healing data generation stage.}
\label{fig:jailbreak-optimization-example}
\Description{}
\end{figure}

\end{document}